\documentclass[12pt]{JHEP3}

\usepackage{axodraw}
\usepackage{cite}
\usepackage{epsfig,multicol}
\usepackage{amsmath,amssymb,mathrsfs}

\newcommand{\ptmiss}{{p \!\! \slash}_T}
 
\newcommand\lsim{\mathrel{\raise.3ex\hbox{$<$\kern-.75em\lower1ex\hbox{$\sim$}}}}
\newcommand\gsim{\mathrel{\raise.3ex\hbox{$>$\kern-.75em\lower1ex\hbox{$\sim$}}}}

\newenvironment{Eqnarray}%
     {\arraycolsep 0.14em\begin{eqnarray}}{\end{eqnarray}}
\newcommand{\beqa}{\begin{Eqnarray}}
\newcommand{\eeqa}{\end{Eqnarray}}
\newcommand{\beq}{\begin{equation}}
\newcommand{\eeq}{\end{equation}}

\def\Eq#1{Eq.~(\ref{#1})}
\def\eq#1{eq.~(\ref{#1})}
\def\eqs#1#2{eqs.~(\ref{#1}) and (\ref{#2})}

\def\half{\hbox{$\frac{1}{2}$}}

\title{Supersymmetric Monojets at the Large Hadron Collider}

\author{Benjamin C.\ Allanach$^{1}$, Sebastian Grab$^2$ and Howard E.\
  Haber$^{2}$
  \\
$^{1}$ DAMTP, CMS, University of Cambridge, Wilberforce Road, Cambridge, CB3
  0WA, UK\\
$^{2}$ Department of Physics and SCIPP, University of California, Santa Cruz,
  CA 95064, USA\\
}
\keywords{Supersymmetry Phenomenology}

\abstract{Supersymmetric monojets may be
  produced at the Large Hadron Collider by the process $q g \rightarrow
  {\widetilde q} {\widetilde \chi}_1^0 \rightarrow q
  {\widetilde \chi}_1^0 {\widetilde \chi}_1^0$, leading to a jet recoiling against missing transverse
  momentum.
  We discuss the feasibility and utility of the supersymmetric
  monojet signal. In particular, we examine the possible precision with
  which one can ascertain the ${\widetilde{\chi}_1^0\widetilde
    q}q$ coupling via the rate for monojet events.  
  Such a coupling contains information on the composition of the
  ${\widetilde \chi}_1^0$ and helps bound dark matter direct
detection cross-sections and the dark matter relic density of the
$ \widetilde{\chi}_1^0$.  It also provides a check of the supersymmetric relation
between gauge couplings and gaugino-quark-squark couplings.}

\preprint{DAMTP-2010-83\\SCIPP-10/16}

\begin{document}

\section{Introduction}

With the first Large Hadron Collider
(LHC) experimental run now underway, searches for physics
beyond the Standard Model will begin to test new theories of TeV-scale
phenomena in parameter regimes that were not accessible to
previous collider facilities.
There are many proposals for theories of TeV-scale
physics beyond the Standard Model.  These include supersymmetry,
technicolor, little Higgs models, extra-dimensions, low-scale quantum gravity,
etc., all of which posit the existence of new classes of
fundamental particles~\cite{pdg}.
In many such models, the lightest member of the new class is
absolutely stable due to the presence of a conserved discrete
symmetry.  As a result, production of the lightest new physics
particle (either directly or at the end of chain of heavier decaying
particles) will lead to missing transverse energy signatures at the LHC.
More generally, signatures from different models of TeV-scale
physics often possess similar features, and distinguishing among
model interpretations may be quite challenging.

In this paper, we assume that after a number of years of LHC running,
a class of new particles has been discovered.  We assume that their
masses and Standard Model quantum numbers will be relatively well determined.
We also assume that we will have some information about the spins of
the new particles.
Under these assumptions, how strong will the case be for a
supersymmetric (SUSY) interpretation of the new physics?

Ideally, one would first try to identify the new particles as
superpartners of Standard Model particles.  Such an attempt will
likely be incomplete, as the entire spectrum of the new physics may
not have been revealed, and the spins of the new particles may not
be reliably known in all cases.  In isolation, the discovery of a
neutral color octet fermion or a color triplet scalar does not
necessarily imply the discovery of the gluino and the squark. Due to
supersymmetry-breaking, the masses of the new particles would not
provide any evidence for a supersymmetric interpretation.  However,
supersymmetry-breaking effects will typically have a small impact
on dimension-four couplings~\cite{Martin:1999hc}.  These couplings
therefore reflect the underlying
supersymmetric structure.   Hence, the measurement of a relation
among couplings could provide very strong evidence for a
supersymmetric interpretation of the new physics.

In the supersymmetric extension of the Standard Model, the couplings
of gluons to squark pairs and gluino pairs
and the couplings of the photon, $W^\pm$ and $Z$ to
squark pairs and slepton pairs
are governed by SU(3)$\times$SU(2)$\times$U(1) gauge
invariance.  In contrast, supersymmetry relates
the gauge couplings of particles (or their supersymmetric partners) to
the Yukawa couplings of gauginos to particle-sparticle pairs.
Similarly, supersymmetry relates the Yukawa couplings of Higgs bosons
to the Yukawa couplings of higgsinos to particle-sparticle pairs.
Thus, the precision measurement of the interaction strengths
of particle-sparticle vertices can provide the smoking gun for a
supersymmetric interpretation of the new physics.

Precision measurements of couplings is usually the domain of $e^+e^-$
colliders.  Indeed, there is a rich program of precision supersymmetry
proposed for an experimental program at the ILC~\cite{Djouadi:2007ik}.
However, until
the ILC becomes a reality (current projections suggest that this is
unlikely during the present decade), we must rely on the LHC running
at high luminosity to provide the necessary data for interpreting the
fundamental nature of new physics discoveries.  Ultimately, a combined
LHC/ILC analysis would yield the most precise tests for the
supersymmetric interpretation of new physics~\cite{Weiglein:2004hn}.

Until now, only a few analyses have been proposed for the LHC.  In
Ref.~\cite{Freitas:2006wd}, same flavor squark pair production at the
LHC was investigated in order to measure the strength of the 
gluino-squark-quark coupling. But this analysis still relied on additional
data from a linear collider. The analysis was refined in
Ref.~\cite{Freitas:2007fd} by taking into account more production
channels of strongly interacting SUSY particles. It was found that a
test of the SUSY relation for the gluino-squark-quark operator
based solely on LHC data might
be possible with a precision better than $10\%$.
A later study~\cite{Brandenburg:2008gd} assuming
300~fb$^{-1}$ of LHC data, with varying degrees of ILC input, found similar
fractional precisions of $3$--$7\%$.  Finally, in
Ref.~\cite{Bornhauser:2009ru}, (left-handed) squark pair production
was considered via color-singlet gaugino exchange in the $t$- and
$u$-channel. It was shown that a measurement of this process, and
therefore a measurement of the wino-squark-quark coupling, might be
possible at the LHC.

We propose the development of a program of precision measurements of
new particle interactions at the LHC in order to provide definitive
evidence in support of a possible supersymmetric interpretation of
TeV-scale physics.  Such a program must necessarily comprise a broad
class of new particle signatures.  Ultimately, a global fit to a
plethora of supersymmetric observables will be required to provide the
maximal coverage of the underlying parameter space.  In this paper, we
take the first step by identifying a particular signature that is
sensitive to the squark-quark-gaugino coupling.

In Section~\ref{Sec:monojets_revisited}, we identify the supersymmetric monojet signal as a
promising arena for measuring the squark-quark-gaugino coupling.  Two
basic supersymmetric scenarios are considered, which
depend on the identity of the lightest supersymmetric particle (LSP).
We contrast the phenomenology of a bino-LSP and a wino-LSP, which
arise in many models of TeV-scale supersymmetry.  In Section~\ref{monojets}, we
outline the process by which one can test the supersymmetric coupling
relations at the LHC.  These procedures are then employed in Sections~\ref{Sec:bino_coupling}
and \ref{Sec:wino_coupling} to measure the squark-quark-gaugino couplings in the case of
the bino-LSP and wino-LSP, respectively.  Finally, we state our
conclusions in Section~\ref{Sec:summary}.  The $p_T$-spectrum of the monojet signal
exhibits a peak at a characteristic value of the transverse momentum
that is approximately given by
\beq \label{Jpeak}
(p_T)_{\rm peak}\simeq \frac{m_{\tilde q}^2-m_{{\tilde \chi}_1^0}^2}{2m_{\tilde q}}\,.
\eeq
This result, which is derived in Appendix A [cf.~\eq{peak}], is
analogous to the Jacobian peak of the electron transverse momentum in
$W$ production and decay.

\clearpage

\section{Supersymmetric Monojets Revisited}
\label{Sec:monojets_revisited}

\subsection{Associated Production of Squarks and Neutralinos}

Supersymmetric monojet signatures have an interesting history. In the early
1980s, CERN $p \bar p$ collisions resulting in single jets (``monojet'')
in association with missing transverse energy in the UA1
experiment~\cite{UA1} were famously
interpreted in terms of supersymmetric particle
production~\cite{Gluck:1983yq,Haber:1984fi,EllisPapers,Barger:1984uz,Barger:1984dt,Allan:1984ds,Tracas:1984aa,Barger:1984nb,Reya:1984ey,Barnett:1985dm,Delduc:1985kh,Gluck:1985iz,DeRujula:1984aw}.
One example of such a process was
squark photino production, followed by decay of the squark
into a quark and photino.
Later, it was calculated that the monojet events would predict more di-jet
missing transverse momentum events than were
observed~\cite{Barnett:1985ci,Reya:1985in}.
Finally, backgrounds (particularly $W$ production,
where $W \rightarrow \tau \nu_\tau$, followed by $\tau \rightarrow \nu_\tau
j
$), and $Zj$ production, where $Z\rightarrow \nu \bar
\nu$~\cite{Ellis:1985sg}, were seen to
adequately account for the monojet events.

Several authors have discussed the ability of LHC experiments to
discover large extra dimensions through the monojets
signal~\cite{Vacavant:2001sd,Benucci:2010ax} predicted by the production of a
hard gluon and
a Kaluza-Klein tower of gravitons. Obvious Standard Model (SM) backgrounds in
the
LHC environment
include di-jet production, where one of the jets is lost, and the $Wj$, $Zj$
backgrounds mentioned above. We wish to {\em measure} the rate of the
dominant SM background (i.e.\ monojet
production via $Zj$) by measuring the cases where $Z \rightarrow \ell^+ \ell^-$  
($\ell=e$, $\mu$) and
using them to predict the background process. Isolated lepton vetoes help
discriminate against $Wj$, and a large missing $p_T$ cut discriminates
against the QCD di-jet background.
More recently, unparticle-jet production, leading to a monojet signature
has also been examined~\cite{Rizzo:2008fp}. The monojet $p_T$ spectrum is
fairly
featureless in either the large extra dimensional case or in the
unparticle-jet case, although the rates are in general different.

Early $\sqrt{s}=7$ TeV LHC
collisions resulting in monojets
recorded by ATLAS~\cite{atlasConf} have already been made public.
The data show no excess
over SM Monte Carlo predictions for 70 nb$^{-1}$ of LHC data, but we find that
the monojet channel
is not yet sensitive to SUSY models that are not already ruled out by other
experiments.\footnote{Note, however, that multijets plus $\ptmiss$ channels are
  already
becoming competitive with published Tevatron
bounds in certain corners of MSSM parameter space~\cite{Alves:2010za}.} Much
more data will be required.

We wish to appraise the ability of the LHC to measure
SUSY monojet production,
assuming the minimal supersymmetric standard model (MSSM) via the
process $qg\to \widetilde q{\widetilde \chi}_1^0$.  At tree-level, there are two
diagrams that contribute, which are exhibited
in Fig.~\ref{fig:feyndiag}.  The tree-level scattering amplitudes for $qg\to\widetilde q{\widetilde \chi}_1^0$
were obtained in Ref.~\cite{matel}.  Strong SUSY-QCD corrections and
the leading logarithmic electroweak corrections at one loop have been
treated in Ref.~\cite{Gounaris:2004cv} (although we note that the one-loop QCD corrections to
$qg\to\widetilde q{\widetilde \chi}_1^0$ do not yet appear in the literature).
\FIGURE{
\begin{picture}(300,100)(0,0)
\Gluon(0,0)(50,50){5}{4}
\ArrowLine(0,100)(50,50)
\CCirc(100,50){5}{Black}{Black}
\Text(90,60)[]{$\lambda$}
\ArrowLine(50,50)(100,50)
\DashLine(100,50)(150,100){5}
\Text(128,65)[]{$\widetilde q$}
\Text(128,35)[]{${\widetilde \chi}_1^0$}
\ArrowLine(100,50)(150,0)
\Gluon(200,0)(250,0){5}{4}
\ArrowLine(200,100)(250,100)
\ArrowLine(250,100)(300,100)
\CCirc(250,100){5}{Black}{Black}
\Text(260,90)[]{$\lambda$}
\DashLine(250,100)(250,0){5}
\DashLine(250,0)(300,0){5}
\Text(285,10)[]{$\widetilde q$}
\Text(285,90)[]{${\widetilde \chi}_1^0$}
\end{picture}
\caption{Tree-level Feynman diagrams leading to $q g \rightarrow {\widetilde q}
  {\widetilde \chi}_1^0$. Monojet signatures result from ${\widetilde q} \rightarrow q
  {\widetilde \chi}^0_1$.\label{fig:feyndiag}}
}

The processes depicted in Fig.~\ref{fig:feyndiag}
do not constitute the best SUSY search channel, since the
production amplitude is proportional to a weak ($\lambda$) times a strong gauge
coupling.
Di-squark,
di-gluino and/or gluino-squark production are expected to have much higher
rates because their amplitudes are proportional to the strong gauge coupling
squared. This is probably why the associated production of squarks and neutralinos
has not been extensively examined
in the literature. On the other hand, if the processes in
Fig.~\ref{fig:feyndiag} could
be identified and the rates measured, an estimate for the coupling $\lambda$
may result. This coupling contains information on the identity of the lightest
neutralino
${\widetilde \chi}_1^0$, which may be dominantly bino, wino, higgsino or a mixture. For
example, in the bino (wino) dominated case, supersymmetry dictates that
$\lambda$ is proportional to $g'$ ($g$), the gauge coupling of the
gauge group U(1)$_Y$ (SU(2)$_L$), respectively (where the proportionality
constant is fixed by supersymmetry). If one knew the constitution
of the neutralino from other measurements, 
then the measurement of the coupling $\lambda$ in associated squark--neutralino
production
would provide a direct test of supersymmetry.\footnote{As emphasized in Section~1,
the measurement of the gluon couplings to gluino pairs or to squark pairs 
does not constitute a test of supersymmetry, as these latter couplings are
governed by QCD.} 

It has long been
postulated that ${\widetilde \chi}_1^0$ constitutes the dark matter of our universe.
Knowing its couplings is a vital ingredient in calculating how much of it is
left as a thermal relic in the universe. For example, neutralinos may
annihilate in the early universe through the process
depicted in Fig.~\ref{fig:DM}a before freezing out. Such a process involves
the same vertex as the
one in our signal.
If ${\widetilde \chi}_1^0$
constitutes the dark matter of the universe, direct detection experiments have
the chance to measure it through it causing nuclear recoils. As depicted in
Fig.~\ref{fig:DM}b, nuclear recoil
mediated via squark exchange  could
contribute a significant part of the
direct detection cross-section~\cite{Belanger:2008sj}. The measurement of
$\lambda$ would bound the direct detection cross-section contribution from
such a channel.
\FIGURE{
\begin{picture}(300,100)(0,0)
\Text(40,100)[]{(a)}
\ArrowLine(50,0)(100,0)
\ArrowLine(100,0)(150,0)
\ArrowLine(50,100)(100,100)
\ArrowLine(100,100)(150,100)
\CCirc(100,100){5}{Black}{Black}
\CCirc(100,0){5}{Black}{Black}
\Text(110,90)[]{$\lambda$}
\Text(110,10)[]{$\lambda$}
\DashLine(100,100)(100,0){5}
\Text(65,10)[]{${\widetilde \chi}_1^0$}
\Text(135,90)[]{$q$}
\Text(110,50)[]{$\widetilde q$}
\Text(65,90)[]{${\widetilde \chi}_1^0$}
\Text(135,10)[]{$q$}
\Text(190,100)[]{(b)}
\ArrowLine(200,0)(250,0)
\ArrowLine(250,0)(300,0)
\ArrowLine(200,100)(250,100)
\ArrowLine(250,100)(300,100)
\CCirc(250,100){5}{Black}{Black}
\CCirc(250,0){5}{Black}{Black}
\Text(260,90)[]{$\lambda$}
\Text(260,10)[]{$\lambda$}
\DashLine(250,100)(250,0){5}
\Text(285,10)[]{${\widetilde \chi}_1^0$}
\Text(285,90)[]{$q$}
\Text(260,50)[]{$\widetilde q$}
\Text(215,90)[]{${\widetilde \chi}_1^0$}
\Text(215,10)[]{$q$}
\end{picture}
\caption{Tree-level Feynman diagrams contributing to (a) dark matter
annihilation and (b) dark matter direct detection.  \label{fig:DM}}
}

There are advantages and disadvantages to the SUSY monojet signature as
compared to monojets resulting from large extra dimensions or unparticle-jet
production.
As we shall illustrate below, SUSY monojet production has features in the
$p_T$ spectrum, which would help to convince us that the distribution is not
merely an incomplete understanding of SM backgrounds.
In particular, the location of the Jacobian peak provides an
indication of the masses of the squark and neutralino (an analytical
approximation is provided in Appendix A).
The disadvantage is that SUSY
backgrounds may be problematic. Other SUSY processes than those in
Fig.~\ref{fig:feyndiag}
leading to the monojet signature include: ${\widetilde \chi}_1^0 {\widetilde \chi}_1^0$ production, with
an initial-state radiated jet, and $\widetilde q \widetilde q$ production, where
each $\widetilde q \rightarrow q {\widetilde \chi}_1^0$, but both quarks are either in the same
direction, or one is lost (for example because it is emitted at high
rapidity). The case where both quarks are in the same direction can be
discriminated against with a maximum jet mass cut.

Various papers have examined SUSY monojets recently, but these are all
different processes from the one we attempt to isolate and would be classified as
SUSY backgrounds by our study.
In Ref.~\cite{Covi:2010au}, a scenario where gluinos and neutralinos are quasi
mass-degenerate were considered. Such a scenario allows a gravitino dark matter
candidate while
being compatible with high reheating temperatures required by thermal
leptogenesis.
This scenario leads to an effective monojet signature via $\widetilde
q \widetilde g$ 
production, where the gluinos decay to soft QCD radiation and quasi-stable
neutralinos, and the squark decays to a jet and a neutralino.
Weak
gaugino pair production plus a jet, leading to monojet
signatures, has recently been examined in Ref.~\cite{Giudice:2010wb} for the
case of quasi-degenerate gauginos, so that the visible products of their
decays are too soft to be detected.
In Ref.~\cite{Carena:2008mj}, SUSY monojet signatures were examined
for a region of the MSSM consistent with baryogenesis and dark matter
constraints. These monojet
signatures originated from stop pair production plus an additional QCD jet,
where the stops decay
invisibly into a soft charm quark and a ${\widetilde \chi}_1^0$.
Monojet searches at the Tevatron have recently been
used to place bounds on dark matter direct detection
rates~\cite{Bai:2010hh} and on indirect dark
matter searches via gamma ray lines~\cite{Goodman:2010qn}. The collider process
investigated in Ref.~\cite{Bai:2010hh} consisted of
initial state radiation in dark matter pair production, which is
classified as a background in the present paper. 
It was found that for very 
light dark matter, below 5 GeV, the inferred Tevatron bounds are stronger than
those from direct detection.\footnote{Although the absence of direct detection
  signals is always subject to a potentially large systematic of unknown
  density of dark matter at the site of the experiment.}
Pair production of dark matter particles in association with one or more
jets at Tevatron and LHC, was also considered in Ref.~\cite{Beltran:2010ww}.
For our study, all processes mentioned above in this paragraph are classified
as SUSY backgrounds. An earlier work~\cite{Klasen:2006kb} included predictions
for monojet signatures from ultra-light gravitinos at the Tevatron and
LHC. This is a different scenario from the one of interest in the
present paper. 

LHC and Tevatron cross sections for total SUSY monojet production were listed in
Ref.~\cite{Dreiner:2009er}, which was primarily concerned with the case of
massless ${\widetilde \chi}_1^0$ particles.  Of course, the
total SUSY monojet production includes our signal process (alongside
SUSY backgrounds). We extend this study in several ways: for example, we
perform a more detailed analysis, finding reasonable cuts to discriminate
signal from background and presenting the kinematics of the events. We also
discuss background subtraction. We focus on the case of massive ${\widetilde \chi}_1^0$
particles and determine with what accuracy $\lambda$ might be measured.

\subsection{The Nature of the LSP}

Supersymmetric collider signals are notoriously complicated and depend
strongly upon the parameter space, even if one restricts oneself to the MSSM.
The monojet signal of interest for this paper depends primarily
on two properties of the supersymmetric model: (i) the precise nature of the LSP
(e.g., the relative contributions of the bino, wino and higgsino components of
the neutralino LSP wave function); and (ii) the branching ratio of the
squark into a quark and the LSP.

In $R$-parity-conserving supersymmetric models, the most likely candidate for the
LSP (excluding the gravitino which is not relevant for this discussion)
is the lightest neutralino ${\widetilde \chi}_1^0$.\footnote{Cosmological and laboratory
constraints rule out other possible candidates such as the sneutrino~\cite{Hebbeker:1999pi}, 
gluino~\cite{gluinoLSP} or charged slepton~\cite{Ellis:1983ew} in almost all possible parameter
regimes.}
In general, ${\widetilde \chi}_1^0$ is a linear combination of bino, wino and higgsino
interaction eigenstates.  The relative contributions of each of these
components depends on the parameters that govern the neutralino mass
matrix.  These include $m_Z$, the higgsino mass parameter $\mu$,
the gaugino mass parameters $M_1$, $M_2$, and the ratio of Higgs
vacuum expectation values $\tan\beta$.
(Approximate formulae for the neutralino masses and mixing matrix in terms of
these parameters can be found in Ref.~\cite{Gunion:1987yh}.)
If the higgsino component were dominant, then the monojet signal proposed in this
paper would not be viable, as the corresponding cross-section for squark-neutralino
production would be suppressed by a light quark Yukawa coupling.  In this case, other
methods must be employed to test for supersymmetric coupling relations.
However, we note that in many (though not all) models of supersymmetry, the parameter
$|\mu|$ is parametrically larger than $M_1$, $M_2$ and $m_Z$, in which case the
higgsino component of the LSP is small.  Henceforth, we will assume that the underlying
SUSY model supports a ${\widetilde \chi}_1^0$ that is dominantly gaugino in nature.

This still leaves the question of the relative contributions of the bino and wino
components of the ${\widetilde \chi}_1^0$ wave function.  In any supersymmetric model with the
unification of tree-level gaugino mass parameters, the ratio of the low-energy
values of the gaugino mass parameters is given by~\cite{Kane:1993td},
\beq \label{gauginomass}
M_1\simeq \frac{5g^{\prime\,2}}{3g^2} M_2\simeq 0.5 M_2\,.
\eeq
Assuming that the tree-level gaugino masses are non-vanishing (and are of
order $m_Z$), then the bino component of the ${\widetilde \chi}_1^0$ wave function is dominant.
This is typical of most mSUGRA models~\cite{Kane:1993td},
but is more general and depends only on the
assumptions outlined above.

Alternatively, it is possible that the tree-level
gaugino masses vanish,
in which case
\eq{gauginomass} (which holds trivially)
is irrelevant.  In this
case, the gaugino mass parameters arise at one-loop.
In particular, a model-independent contribution
to the gaugino mass is present whose origin can be traced to the super-conformal
(super-Weyl) anomaly, which is common to all supergravity
models \cite{anomalymed}.  This contribution is dominant in models of
anomaly-mediated supersymmetry breaking (AMSB).
\Eq{gauginomass} is then replaced (in the one-loop approximation) by
\beq \label{eqanom}
M_i\simeq {b_i g_i^2\over 16\pi^2}m_{3/2}\,,
\eeq
where $m_{3/2}$ is the gravitino mass (typically assumed to be on the order of 1~TeV),
and the $b_i$ are the coefficients of the MSSM gauge beta-functions
corresponding to the corresponding U(1)$_Y$, SU(2)$_L$, and SU(3) gauge groups:
$(b_1,b_2,b_3)= (\tfrac{33}{5},1,-3)$.
\Eq{eqanom} yields $M_1\simeq 2.8M_2$, which implies
(under the assumption that $|\mu|$ is somewhat larger than $M_2$) that
the lightest chargino pair and neutralino comprise a
nearly mass-degenerate triplet of winos
over most of the MSSM parameter space.  Typically, the corresponding neutralino
is the LSP, whereas the wino is the next-to-lightest supersymmetric particle (NLSP).\footnote{We
ignore the gravitino, which could be lighter than the neutralino.  
Nevertheless, in almost all cases, the neutralino behaves as if it is the LSP.
That is, the neutralino will be sufficiently long-lived so
that it will always escape the collider detector before it decays.}

The squarks that are produced in Fig.~\ref{fig:feyndiag} are primarily first-generation
squarks, either $\widetilde q_L$ or $\widetilde q_R$ (as the mixing between these
states is negligible).  In the case where the LSP is dominantly bino-like,
then the branching ratio ${\rm B}(\widetilde q_R\to q{\widetilde \chi}_1^0)\simeq 100\%$.  In
contrast, the dominant channel for $\widetilde q_L$ decay is into the heavier
wino-like neutralino which subsequently decays into the LSP.  Thus,
the production of $\widetilde q_R$ in association with the LSP is more likely to produce
a monojet.  In the case where the LSP is dominantly wino-like,
the reverse is true, and the production of $\widetilde q_L$ in association with the LSP
is more likely to produce a monojet.

In this paper we shall examine several scenarios.
The first scenario considered will be an optimistic mSUGRA
point, which will illustrate the effect of the SUSY backgrounds, making a
coupling extraction from LHC data difficult. Next, we shall
assume a wino
dominated neutralino, where signal cross-sections are higher and easier to
measure over the background.  In this case, the coupling extraction from LHC data
is somewhat easier.  If the lightest neutralino is light and found to
be wino dominated, then it does {\em not}
constitute a significant portion of a thermal dark matter relic, since winos
annihilate too efficiently in the early universe~\cite{Gherghetta:1999sw}.

\section{Testing the Supersymmetric Coupling Relations with Monojets}
\label{monojets}

\subsection{The Signal Process}
\label{signal_process}

We define our parton-level signal processes
in the following way:
\begin{itemize}
\item For a bino LSP: $g+u \rightarrow \widetilde{\chi}_1^0 + \widetilde{u}_R$, $g+d
  \rightarrow \widetilde{\chi}_1^0 + \widetilde{d}_R$
as displayed in Fig.~\ref{fig:feyndiag} followed by ${\widetilde u}_R \rightarrow
u {\widetilde \chi}_1^0$ or ${\widetilde d}_R \rightarrow d {\widetilde \chi}_1^0$.
Here, $\lambda=g'$, the U(1)$_Y$ gauge coupling of the SM.
The charge-conjugated processes are also included.
\item For a wino LSP: $g+u \rightarrow \widetilde{\chi}_1^0 + \widetilde{u}_L$, 
$g+d \rightarrow \widetilde{\chi}_1^0 + \widetilde{d}_L$,
$g+u \rightarrow \widetilde{\chi}_1^+ + \widetilde{d}_L$, $g+d \rightarrow
  \widetilde{\chi}_1^- + \widetilde{u}_L$, where ${\widetilde u}_L\rightarrow u {\widetilde \chi}_1^0$/$d{\widetilde \chi}_1^+$,
${\widetilde d}_L\rightarrow d {\widetilde \chi}_1^0$/$u{\widetilde \chi}_1^-$ and $\widetilde{\chi}_1^+ \rightarrow
  S+{\widetilde \chi}_1^0$, where $S$ is either QCD radiation too soft to be identified as
  a jet, or a lepton so soft that it passes the lepton veto.
For wino LSPs, the mass splitting between ${\widetilde \chi}_1^+$ and ${\widetilde \chi}_1^0$ is $\sim
200$ MeV, so the dominant decay ${\widetilde \chi}_1^+ \rightarrow {\widetilde \chi}_1^0 \pi^+$
typically results in a soft pion that is too soft to be measured by the LHC
experiments.
Both the $\widetilde{\chi}_1^0$ and
$ \widetilde{\chi}_1^+$ are wino-like
and are related via a SU(2)$_L$ transformation, so
the production process amplitudes are constrained by the MSSM to be
proportional to $\lambda$, equal to the SU(2)$_L$ gauge coupling $g$. 
The charge-conjugated processes are also included.
\end{itemize}
In principle, processes involving heavier non-valence quarks
like $g+c \rightarrow \widetilde{\chi}_1^0 + \widetilde{c}$ also contribute to our
signal and SUSY backgrounds. We have only included initial states with a
$g,u,\bar u, d$ or $\bar d$, since processes
with quarks beyond the first generation are negligible.
Since we assume that ${\widetilde \chi}_1^0$ is stable, it leaves no direct trace in the
detector, and so the signal consists of a jet recoiling against apparently
missing transverse momentum of magnitude $\ptmiss$. In general, $\ptmiss$ is
measured to be
different to the transverse momentum of the jet $p_T(j)$ because of
measurement errors and also because of soft QCD radiation, which may not be
included in the jet, but measured in the calorimeter nonetheless.


\subsection{Major Backgrounds and Basic Cuts}
\label{Sec:backgrounds}

As we have seen in the last section, our signal consists of a hard jet recoiling
against missing energy. Therefore, the major SM backgrounds are \cite{Vacavant:2001sd,Vacavant:2000wz}
\begin{itemize}
\item $Z(\rightarrow \nu \, \bar \nu)$+jet.  $Z$ production in association with a hard
jet, where the $Z$ decays into a pair of neutrinos.
\item $W(\rightarrow \tau \nu)$+jet. $W$ plus jet production, where the $W$
  decays into a tau and a neutrino, and the tau is either not detected, or
  lost in the jet.
\item $W(\rightarrow e/\mu \, \nu)$+jet. $W$ plus jet production, where the
  $W$ decays into an
  electron or a muon and a neutrino and the electron/muon is undetected or
  lost inside the jet.
\item QCD jet production together with mismeasurement
of the energy deposited in the detector. One could produce di-jets, for
instance, and one of the jets could be lost in the detector (or its energy
mismeasured so that it fluctuate below the transverse momentum required to
identify the jet).
\end{itemize}
In principle, di-vector boson production $VV'$, with
$V,V'=W,Z$ is an
additional source for the SM backgrounds if one vector decays into a neutrino
and the
other into jets. However, the cross section is much smaller
than $Z/W+$jet and $VV'$ production can be safely neglected in our analysis
\cite{Rizzo:2008fp}.

In order to reject most of the $W$+jet background, we employ
a lepton veto.
We veto events with an isolated electron or muon with $p_T>5$ GeV and with
$|\eta|<2.5$. The isolation
criterion demands less than 10 GeV of additional energy in a cone of radius
$\Delta R=
\sqrt{\Delta \phi^2 + \Delta \eta^2}= 0.2$ around the lepton momentum, where
$\phi$ and $\eta$ are the azimuthal angle and pseudorapidity,
respectively.

After the lepton veto, $Z(\rightarrow \nu \bar \nu)$+jet remains the most
important irreducible background.
Fortunately, this background can be directly derived from data itself by
measuring  $Z(\rightarrow e^+e^-/\mu^+\mu^-)$+jet and relating both processes
with the help of the measured $Z$ branching ratios. Systematic uncertainties
are in this case significantly
reduced \cite{Aad:2009wy}. However, this comes with the cost of higher
statistical uncertainties,
because the $Z(\rightarrow \nu \bar \nu)$+jet cross section is roughly three
times larger than the  $Z(\rightarrow e^+e^-/\mu^+\mu^-)$+jet cross section.
$\gamma$+jet cross-sections at high $p_T$ could be used in the future to
estimate the $Z(\rightarrow \nu \bar \nu)$+jet cross section with better
statistics.

We will use two statistical estimators;
cf.~Refs.~\cite{Vacavant:2001sd,Vacavant:2000wz}.
The first {\it optimistic} estimator takes only the statistical fluctuations
from $Z(\rightarrow \nu \bar \nu)$+jet into account. In this case the
significance
is given by $S/\sqrt{B}$ with $S$ ($B$) the number of expected signal (SM
background)
events. This case would apply when the background Monte-Carlo is so well tuned
and tested with LHC data that its output may be fully trusted.
We also employ a {\it conservative} estimator, where the statistical
fluctuations
are dominated by the $Z(\rightarrow e^+e^-/\mu^+\mu^-)$+jet calibration sample.
According to Refs.~\cite{Vacavant:2001sd,Vacavant:2000wz}, this case
corresponds to a significance of $S/\sqrt{7\,B}$.\footnote{In 
Refs.~\cite{Vacavant:2001sd,Vacavant:2000wz} the transverse
  momentum  of the monojet was required to lie above 1~TeV. In contrast, we
  will use less hard  cuts for which the $Z(\rightarrow
  e^+e^-/\mu^+\mu^-)$+jet acceptance is slightly
better; see Ref.~\cite{Vacavant:2000wz}. However, we will always use
$S/\sqrt{7\,B}$
as a conservative estimator.} The conservative estimate would be used in the
case that one does not trust at all the Monte-Carlo background calculation,
and instead measures the $Z+$jet background from LHC data.
We expect the true (i.e.~measurable) significance to lie between our two
estimators after the
Monte Carlo
estimates \cite{Aad:2009wy} have been properly tuned to LHC data, including
$W$+jet/$\gamma$+jet. The modeling of systematic error on the measurements of
SM backgrounds is an experimental question, and as such is beyond the scope of
this paper.

After cuts, 
once the large $Z(\rightarrow \nu \bar \nu)+$jet background is subtracted, we
will still need to subtract the $W+$jet background, which is larger than our
signal. This can be measured in 
the case where the lepton from the $W$ is visible, by extrapolating into 
the region where the lepton is invisible, either because it is lost in the jet
(in which case one could extrapolate in $\Delta R$ between jet and
lepton), or because the lepton is missed. 

An estimate of the QCD background can be found
by full detector simulation \cite{Aad:2009wy}.
The quantity of interest for estimating these backgrounds is the jet energy
response
function (JERF) $R$,
equal to the ratio of the measured jet energy to the true one.
We fit the JERF to the full
detector simulation results in  \cite{Aad:2009wy} in the same spirit as
Ref.~\cite{Barr:2009wu}.
In our analysis, $R$ is
used to scale all of the components of the jet four-momentum.
$R$ is well fit by a probability distribution function
\begin{equation}
p(R)=\frac{0.99}{\sqrt{2 \pi \sigma^2}}\ e^{\frac{(1-R)^2}{2 \sigma^2}} + 0.01\
A\ e^{7.32R}\
\Theta(0.9-R),
\end{equation}
where $\Theta$ is the Heaviside step function,
$A$ is an unimportant normalization constant ($\approx 0.01$), and
\begin{equation}
\sigma = \frac{0.6}{\sqrt{E/{\rm GeV}}} \oplus 0.03
\end{equation}
is the resolution of the Gaussian part of the JERF, and $\oplus$ denotes the
fact that the two terms are added in quadrature.  The second term in $p(R)$ reflects the 1$\%$ probability that
a significant portion of the jet will go unmeasured due to cracks in the
detector and other
effects. This exponential tail is responsible for the QCD
background to our SUSY monojet signal, and in order to increase statistics on
the sample, we impose that
either the hardest or second hardest jet in the sample be in the Gaussian
tail, taking into account the additional factor of 100 needed to calculate the
cross-section after cuts. Other jets are all drawn from $p(R)$.
QCD backgrounds to monojets will be extrapolated
from data, for example from di-jet, $\gamma+$jet and ``Mercedes'' type 3-jet
events~\cite{Aad:2009wy}, although we note that these techniques may not
address more conspiratorial backgrounds, which could require the use of tracking
information. QCD backgrounds will then be subtracted from event samples,
leaving statistical fluctuations only once the systematics have been dealt
with. The JERF was only applied to QCD backgrounds, not to the other samples,
since it should only have a small effect upon them.  We note that {\tt
  Herwig++} includes $b$ backgrounds within this QCD sample, so the case of
the production of
$b \bar b$, where one of the bottom quarks' momenta primarily goes into a
neutrino
resulting in a monojet signature, is included in our estimate.

A further background source is supersymmetry itself.  For example,
squark pair production and the subsequent decay of the squarks into
the $\widetilde{\chi}_1^0$ LSP and a quark is a possible SUSY background.
If the two jets from the squark decays overlap they can appear as a
single monojet. Initial-state radiation on ${\widetilde \chi}_1^0 {\widetilde
  \chi}_1^0$ production also provides a SUSY background, although this
is usually rather small.  Note that the background of squark pair
production followed by the decay into two jets plus two
${\widetilde\chi}_1^0$, where one of the jets is in the tail of the JERF,
is {\em not} in our SUSY background sample. We expect this background
to be smaller than the QCD background, which is already small.

In the case of a bino-like LSP, $\widetilde\chi_2^0+\widetilde q_L$
production followed by $\widetilde\chi_2^0\to\widetilde\chi_1^0+\nu
\overline\nu$ and $\widetilde q_L\to \widetilde\chi_1^0+q$ would also
produce monojet events.  The cross-section for
 $\widetilde\chi_2^0+\widetilde q_L$ is enhanced by
 $Kg^2/g^{\prime\,2}\simeq 2$ over  $\widetilde\chi_1^0+\widetilde
 q_R$ production (where the kinematical factor $K$ provides a
 suppression due to the heavier $\widetilde\chi_2^0$ mass).
However, due to the branching ratio factors for the final state decays,
the number of $\widetilde\chi_2^0+\widetilde q_L$ events that are
observed as monojets is significantly smaller than our signal and can
be neglected.

In the case of a (not too heavy; see Section~\ref{Sec:parameter_scan}) wino-like LSP,
the dominant SUSY background is wino pair production
plus a jet from initial state radiation. The cross section is quite large
because two winos,
$ \widetilde{W} \widetilde {W}$, can be produced via
a Drell-Yan like process \cite{Dawson:1983fw,Baer:1986vf},
 i.e.~$PP \rightarrow \gamma^*/Z^* \rightarrow \widetilde{W} \widetilde{W}$.
Here, $\gamma^*$ ($Z^*$) denotes a virtual $\gamma$ ($Z$).
In the wino-LSP scenario, both $\widetilde{W}$ means either
${\widetilde \chi}_1^\pm$ or ${\widetilde \chi}_1^0$, both of which are approximately
winos.
Because the ${\widetilde \chi}_1^\pm$ is
quasi-degenerate in mass with the ${\widetilde \chi}_1^0$, it decays into a soft pion and
the ${\widetilde \chi}_1^0$. The pion is typically too soft to be
detected \cite{Gherghetta:1999sw} unless special analysis techniques are
used~\cite{Barr:2002ex} and so the ${\widetilde \chi}_1^\pm$
is effectively invisible to the detector.
The additional jet is then produced via initial state radiation.


\TABLE{
\begin{tabular}{|c|c|c|}
 \hline
 sample & simulated events & comments \\
 \hline
 $Z(\rightarrow \nu \bar \nu)$+jet & $2\,140\,000$ & $p_T({\rm jet})>70$ GeV \\
 $W(\rightarrow e \nu)$+jet & $2\,960\,000$ & $p_T({\rm jet})>70$ GeV  \\
 $W(\rightarrow \mu \nu)$+jet & $2\,960\,000$ & $p_T({\rm jet})>70$ GeV  \\
 $W(\rightarrow \tau \nu)$+jet & $2\,960\,000$ & $p_T({\rm jet})>70$ GeV  \\
QCD & $18\,650\,000$  &  $p_T({\rm jet})>30$ GeV  \\
\hline
 mSUGRA & $3\,280\,000$ & Included $10\,300$ signal events. \\
 mAMSB & $2\,940\,000$ & Included $46\,700$ signal events. \\
 others & $100\,000$ & Number of signal events varies.\\
 \hline 
 \end{tabular}
\\
\caption{\label{Tab:MC_samples} 
Monte Carlo samples of SM backgrounds and SUSY events 
used in our analysis.  All simulated events were generated with
\texttt{Herwig++2.4.2}
for $pp$ collisions at a center-of-mass energy of $14$~TeV.
The $Z$+jet and $W$+jet samples correspond to an integrated
luminosity of 5~fb$^{-1}$.  The QCD sample corresponds to 30~nb$^{-1}$,
the mSUGRA sample corresponds to 20~fb$^{-1}$ and
the mAMSB sample corresponds to 100~fb$^{-1}$. 
The cut on $p_T$(jet) is performed at the parton level.     
The mSUGRA [mAMSB] sample corresponds
to sparticle pair production assuming the benchmark scenario
\eq{benchmark_mSUGRA} [\eq{mAMSB_param}]; the number of SUSY
events that correspond to our signal monojet events is indicated in 
the third column of the Table for these two rows.  The scenarios denoted
by {\it others} are those of the parameter scan of 
Section~\ref{Sec:parameter_scan}; the corresponding integrated
luminosities have been adjusted for each scan point such that
100\,000 events for each point are generated.
 }}

We have employed {\tt
  Herwig++2.4.2}~\cite{Lonnblad:1998cq,Bahr:2008pv,Bahr:2008dy}\footnote{We 
  have used a modified version of {\tt Herwig++2.4.2} which is also
able to deal with negative (Majorana) gluino masses; see the following link
for details \cite{gluino}.}
to simulate the signal and the backgrounds at tree level
in $pp$ collisions at a center-of-mass energy of 14~TeV.
An overview of the
simulated signal and background samples is given in Table~\ref{Tab:MC_samples}.
We have included the pair
production and the two and three-body decay of {\it all} SUSY particles.
All events from this sample that yield the correct monojet
topology that are not our signal process
are classified as SUSY background. The $Z$+jet/$W+$jet backgrounds are also
obtained with {\tt   Herwig++}.
The {\tt Herwig++} output was analyzed using {\tt HepMC-2.04.02} \cite{Dobbs:2001ck}
and {\tt ROOT} \cite{Brun:1997pa,Canal:2010zz}. Jets were reconstructed with the help of
{\tt fastjet-2.4.1} \cite{fastjet}. We employed the anti-$k_t$ jet algorithm with
$R=0.7$
\cite{Cacciari:2008gp}. Only jets with $p_T>30$ GeV are used in our analysis
in order to be less sensitive to underlying event modeling. 
In Table~\ref{Tab:MC_samples},
we list the total number of simulated events for each sample. In the bottom
half of the table, the total number of SUSY pair production events is listed
first, with the number in the subset corresponding to our signal events listed
under ``comments''.

\section{Measurement of the $\widetilde{\chi}_1^0\widetilde{q}q$ Coupling for a
Bino LSP}
\label{Sec:bino_coupling}

In this section we will show that it is possible to estimate the $\widetilde{\chi}_1^0\widetilde{q}_R q$
coupling for the case of a bino-like $\widetilde{\chi}_1^0$ LSP from data. We choose
an mSUGRA benchmark scenario for investigation.

\subsection{Benchmark Scenario}
\label{Sec:bechmark_mSUGRA}

Our benchmark scenario is a light mSUGRA scenario \cite{msugramodel} with a bino-like $\widetilde{\chi}_1^0$
LSP. It is described by the parameters
\begin{equation}
M_0=220\, {\rm GeV}, \, M_{1/2}=180\,{\rm GeV}, \, A_0=-500\,{\rm GeV}, \, \tan \beta=20, \, {\rm sgn}(\mu)=+1 \, .
\label{benchmark_mSUGRA}
\end{equation}
For this scenario, the $\widetilde{\chi}_1^0\widetilde{q}_R q$ coupling is given by
$\lambda=0.99 g'$, where the deviation from $\lambda=g'$ is due
to the small admixture of wino and
higgsino eigenstates in the LSP. 
All spectra are calculated with {\tt SOFTSUSY3.0.13}~\cite{Allanach:2001kg} 
and fed into the event generator via
the SUSY Les Houches Accord~\cite{Skands:2003cj}.
The sparticle masses are given in Table~\ref{mSUGRA_scenario}.

\TABLE{
\begin{tabular}{|cc|cc|}
 \hline
 sparticle & mass [GeV] & sparticle & mass [GeV] \\
 \hline
 $\widetilde{\chi}_1^0$ & 70.2 &  $\widetilde{\chi}_4^0$ & 365 \\
 $\widetilde{\chi}_1^+$ & 132 &  $\widetilde{\chi}_2^+$ & 370 \\
 $\widetilde{\chi}_2^0$ & 133 &  $\widetilde{b}_1$ & 378 \\
 $\widetilde{\tau}_1$ & 189 & $\widetilde{b}_2$ & 443 \\
 $\widetilde{t}_1$ & 226 &  $\widetilde{u}_R/\widetilde{c}_R$ & 454 \\
 $\widetilde{\nu}_\tau$ & 230 & $\widetilde{d}_R/\widetilde{s}_R$ & 455\\
 $\widetilde{e}_R/\widetilde{\mu}_R$ & 234 &  $\widetilde{g}$ & 456\\
 $\widetilde{\nu}_e/\widetilde{\nu}_\mu$ & 242 & $\widetilde{u}_L/\widetilde{c}_L$ & 463\\
 $\widetilde{e}_L/\widetilde{\mu}_L$ & 255 &   $\widetilde{d}_L/\widetilde{s}_L$ & 470 \\
 $\widetilde{\tau}_2$ & 259 &  $\widetilde{t}_2$ & 477 \\
 $\widetilde{\chi}_3^0$ & 359 &  &  \\
 \hline
 \end{tabular}
\caption{\label{mSUGRA_scenario} Sparticle mass spectrum of the mSUGRA benchmark
scenario given by eq.~(\ref{benchmark_mSUGRA}). The sparticles are ordered by
their mass.
}
}

This scenario has a relatively light SUSY mass spectrum and a 
total signal cross-section of $\sigma(pp \rightarrow {\widetilde q}_R {\widetilde
  \chi}_1^0)=520$ fb.
For spectra a little heavier,
the monojet cross section is too small and could not be
seen above the SM backgrounds. Note that the mSUGRA scenario,
eq.~(\ref{benchmark_mSUGRA}),
lies at the edge of the region excluded by the Tevatron, but is still allowed
\cite{Aaltonen:2008rv,Abazov:2007ww}.

\subsection{Event Numbers and Cuts}
\label{Sec:cuts_mSUGRA}

In this section, we will develop a set of cuts that allow a measurement of the
$ \widetilde{\chi}_1^0 \widetilde{q}_R q$ coupling assuming our mSUGRA benchmark
scenario, eq.~(\ref{benchmark_mSUGRA}). On one hand,
the number of signal events needs to significantly exceed the statistical
fluctuations of the SM backgrounds. On the other hand, we desire a good signal
to SUSY background ratio
if we want to measure $\lambda$ to a high precision.

\TABLE[b]{
\begin{tabular}{|l|ccc|c|}
 \hline
 cut & all SM &  SUSY bkg. & signal & $S/\sqrt{B}$ \\
 \hline
 trigger & $1.14 \times 10^8$ & $2.91 \times 10^7$ & $130\,000$ & - \\
 lepton veto & $7.57 \times 10^7$ & $1.76 \times 10^7$ & $130\,000$ & - \\
 number(jets)=1 & $3.35 \times 10^7$ & $55\,900$ & $35\,100$ & 6.1 (2.3) \\
 $\ptmiss > 180$ GeV & $3.28 \times 10^6$ & $32\,300$ & $22\,300$ &  12 (4.7) \\
 $m$(jet) $<$ 70 GeV & $3.00 \times 10^6$ & $12\,100$ & $20\,100$ &  12 (4.4) \\
 \hline
 tau veto & $2.75 \times 10^6$ & $9\,950$ & $20\,000$ & 12 (4.6) \\
 $b$-jet veto & $2.66 \times 10^6$ & $9\,290$ & $20\,000$ & 12 (4.6) \\
 \hline
  \end{tabular}
\caption{\label{cutflow_mSUGRA} Cut flow for the mSUGRA benchmark scenario in
  eq.~(\ref{benchmark_mSUGRA}).
We present the cuts in the first column and the number of SM, SUSY background and signal events
in the second, third and fourth column, respectively.
We also show in the fifth column the resulting
significance for the monojet signal, i.e.~$B$
corresponds to the number of SM background events and $S$ is the
number of signal events. The significances in brackets are our conservative
estimate, i.e.~$S/\sqrt{7\,B}$. We have assumed
an integrated luminosity of 300~${\rm fb}^{-1}$ at $\sqrt{s}=14$ TeV.
Note that in addition to the $\ptmiss$ cut, the same cut has been applied
on the jet-$p_T$.}
}

In Table~\ref{cutflow_mSUGRA}, we display the total number of SM background
events (second column),
the number of SUSY background events (third column) and the number of signal events (fourth column)
for different sets of cuts (first column). We also show in the fifth column the resulting
significance for the monojet signal, i.e.~$B$
corresponds to the number of SM background events and $S$ is the
number of signal events.  Note that
$B$~refers to the SM background only, but the SM contribution completely dominates the
background.
We assumed an integrated luminosity of 300 ${\rm fb}^{-1}$ at $\sqrt{s}=14$ TeV.
We cannot give meaningful numbers for the significances in the first two rows
because SM backgrounds additional to those of Section~\ref{Sec:backgrounds} would
need to be included.

As a first cut (denoted by {\it trigger}), we demand at least one jet with
transverse momentum, $p_T$(jet),
larger than 100 GeV. In addition, the amount of missing transverse momentum,
${\ptmiss}$, must also exceed 100 GeV.
These cuts correspond to the planned one-jet plus missing energy
trigger used by the ATLAS
collaboration \cite{Hauser:2004nd} for the 14 TeV run. Therefore, all events
in the first row of Table~\ref{cutflow_mSUGRA}
will be recorded. Note that CMS plans to use harder cuts
\cite{Giordano:2006pz}.
We can see in Table~\ref{cutflow_mSUGRA} that two orders of magnitude more
SUSY background events will pass the trigger than signal events.

The SUSY background is reduced by roughly a factor of two after we apply (in addition to the trigger cut)
a veto on isolated electrons and muons; see Section~\ref{Sec:backgrounds}. At the same time, the number
of signal events is nearly unaffected. The SM backgrounds are also reduced by the lepton veto,
because most of the $W$+jet background events will not pass this cut. However, the
$Z(\rightarrow \nu \bar \nu)$+jet background stays nearly the same.

To reduce the SUSY background further, we also apply a veto on a second jet if its $p_T$ is
larger than 30 GeV and $|\eta|<5$. This corresponds to a veto on a second jet, because we
only count jets above 30 GeV, cf.~Section~\ref{Sec:backgrounds}.
The number of SUSY background events now has the same order of magnitude as
the signal,
namely $\mathcal{O}(10^4)$ events. Although the signal possesses no second jet at parton
level, we might produce one due to initial and final state radiation, as is
borne out by the third row of Table~\ref{cutflow_mSUGRA}.  In particular,
a veto on a second jet also reduces the number of signal events by roughly a factor of four.

At this stage we are able to give some reliable numbers for signal
significances, assuming that backgrounds can be reliably constrained and
subtracted.
The numbers in brackets correspond
to our conservative estimate as described in Section~\ref{Sec:backgrounds}, 
i.e.~we assume that the dominant $Z(\rightarrow \nu \bar \nu)$+jet background is
estimated  purely
from a measurement of $Z(\rightarrow e^+e^-/\mu^+\mu^-)$+jet. We observe an
optimistic (conservative) significance for the monojet signal of 6.1 (2.3).

\FIGURE[t]{
\unitlength=1cm
\begin{picture}(9,8.5)
\put(-1.3,0.34){
\epsfig{file=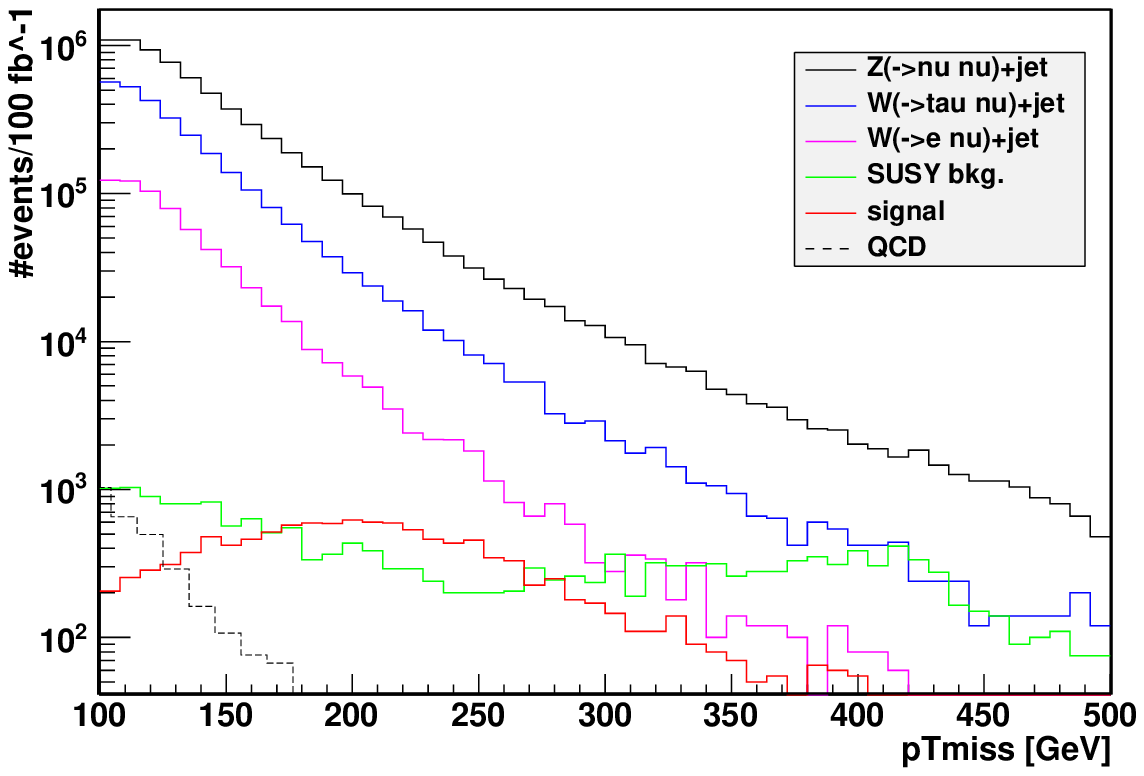,width=11.5cm}
}
\end{picture}
\caption{Missing transverse momentum distribution for the $Z(\rightarrow \nu \bar \nu)$+jet
(black histogram), $W(\rightarrow \tau \nu)$+jet (blue histogram), $W(\rightarrow e \nu)$+jet
(magenta histogram), the SUSY background (green histogram), the QCD background
  (black dashed histogram) and the signal process (red histogram).
The first three cuts of Table~\ref{cutflow_mSUGRA} have been applied. The
number of events correspond
to an integrated luminosity of 100 ${\rm fb}^{-1}$ at $\sqrt{s}=14$ TeV. We
assume the mSUGRA scenario of eq.~(\ref{benchmark_mSUGRA}).
}
\label{Fig:pts_mSUGRA}
}
\FIGURE[t]{
\epsfig{file=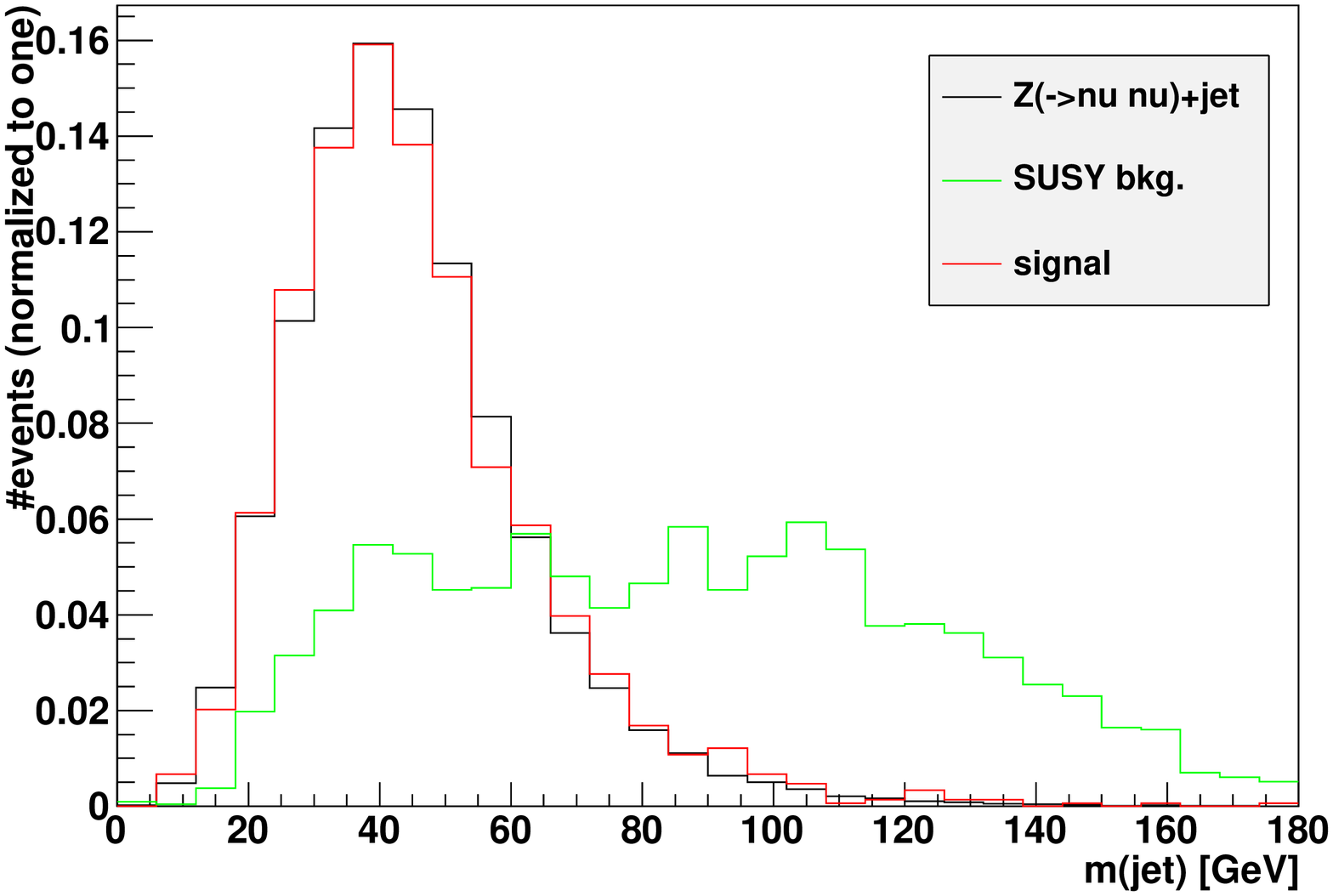,width=7.7cm,angle=270}
\caption{Invariant mass distribution of the hardest jet for the
$Z(\rightarrow \nu \bar \nu)$+jet background (black histogram), the SUSY background (green histogram) and
the signal process (red histogram). The first four cuts of Table~\ref{cutflow_mSUGRA}
have been applied. The distributions are normalized to one. For the signal, we assume the mSUGRA scenario
of eq.~(\ref{benchmark_mSUGRA}). The $W$+jet and QCD backgrounds have a distribution almost
indistinguishable from that of $Z(\rightarrow \nu \bar \nu)$+jet.}
\label{Fig:m_mSUGRA}
}%

In order to increase the signal to SM background ratio we make use of
the ${\ptmiss}$ and jet-$p_T$ distribution of our monojet signal. They possess a Jacobian peak, where
the position depends on the ${\widetilde q}_R$ and $\widetilde{\chi}_1^0$ masses (see Appendix A).
We present the ${\ptmiss}$ distributions in Fig.~\ref{Fig:pts_mSUGRA}. The
Jacobian peak of our
monojet signal (red histogram) lies around 180 GeV. At the same time
the distributions of the SM backgrounds (black, blue, magenta and dashed black histograms)
fall off exponentially.
An additional lower bound on ${\ptmiss}$ {\it and} the monojet $p_T$ of 180
GeV, raises the optimistic (conservative)
signal versus SM background significance to 12 (4.7).
The required level of cancellation is high: around 1 in 100 for a high $\ptmiss$
cut. This means that
the experimental systematic on the dominant background also needs to be at
the level of $\lsim 1\%$, a challenging (but not obviously impossible)
proposition. 
The experimental systematic on the subdominant $W+$jet
background subtraction would need to be at the level of a few percent. 

Our goal is to reconstruct the $\widetilde{\chi}_1^0 \widetilde{q}_R q$ coupling as
precisely as possible. In order to accomplish this goal, 
we need to reduce the SUSY background
further while leaving the number of signal events unchanged (in order to have a visible signal).
For this purpose, we have examined the invariant mass of the hardest
jet,  $m$(jet), which is presented in Fig.~\ref{Fig:m_mSUGRA}
for the signal (red histogram), the SUSY background (green histogram) and the
$Z$+jet
background (black histogram). Each histogram is separately normalized so that
the total number of events is one. The first four cuts of Table~\ref{cutflow_mSUGRA} are applied.
Note that the shape of $m$(jet) for $W$+jet and QCD backgrounds
follows those of $Z$+jet.

We first observe that the distribution of the $Z$+jet background looks very
similar to
the signal distribution. This is expected; the $m$(jet) distribution for these
two cases is what is predicted by an approximately massless initial parton
(which ideally would peak at zero mass) when one applies lower $p_T$(jet)
cuts, cutting out the very low mass region.
Therefore, a cut on $m$(jet) cannot increase the
signal significance of the signal over SM background.
However, the SUSY background distribution has a very different shape. It is
relatively flat compared to the
signal and has its maximum at a larger $m$(jet) value. The (on average) larger
jet invariant  mass of the SUSY background stems mainly from events, where a
pair of squarks is produced and where the jets from squark decays go roughly
in  the
same direction and are identified as only one jet.
Relative to the cuts listed in Table~\ref{cutflow_mSUGRA}, 
imposing the requirement that $m$(jet) is less
than 70 GeV reduces the SUSY background by roughly a factor of three and
leaves the number
of signal events nearly unchanged.\footnote{For the
mSUGRA benchmark point, the number of surviving monojet events 
after the cut on $m$(jet) arising from
$\widetilde\chi_2^0+\widetilde q_L$ production 
(which is included in the SUSY background in Fig.~\ref{Fig:m_mSUGRA})
is about $5\%$ of our signal events.}
Note that the cut on $m$(jet)
mostly  suppresses
the high-$\ptmiss$ SUSY background events in Fig.~\ref{Fig:pts_mSUGRA}, 
i.e.~$\ptmiss \gtrsim 300$ GeV.
In the next section, we will employ these cuts,
i.e.~the first five cuts in Table~\ref{cutflow_mSUGRA}, for the coupling reconstruction.

Finally, in the last two rows of Table~\ref{cutflow_mSUGRA} we show the effects
of a tau-lepton veto and a subsequent $b$-jet veto. In both cases
we assumed an identification (ID) efficiency of 100\%. We regard a tau as identified
if it decays hadronically and has $p_T>15$ GeV and $|\eta|<2.5$.
We observe that neither veto helps much, and so their effect would be even more
reduced if realistic ID efficiencies were assumed. We see that the SUSY background
would be reduced by $18\%$. However, the ID efficiencies are doubtless too optimistic
and in reality the suppression of the SUSY background would lie below $11\%$
\cite{Aad:2009wy,Desch:2010gi}. Thus, we will not employ the tau and $b$-jet vetoes here.

\subsection{Coupling Reconstruction}
\label{Sec:coupling_reco_mSUGRA}

We now show to which precision the $\widetilde{\chi}_1^0 \widetilde{q}_R q$ coupling
$\lambda$
of the mSUGRA benchmark scenario of Section~\ref{Sec:bechmark_mSUGRA} can be reconstructed
with an integrated luminosity of $300\, {\rm fb}^{-1}$ at $\sqrt{s}=14$ TeV.
We employ the cuts developed in Section~\ref{Sec:cuts_mSUGRA} (without the tau and $b$-jet veto).
Since the signal monojet cross section is proportional to $\lambda^2$, the relative error on $\lambda$
will be roughly one half of the relative error on the monojet production cross section. Note
that we reconstruct $\lambda$ under the assumption that the SUSY backgrounds
come from our mSUGRA point.

\TABLE{
\begin{tabular}{|l|cc|}
 \hline
 error & $\Delta \sigma_{\rm mono}/\sigma_{\rm mono}$ &  $\Delta \lambda/\lambda$\\
 \hline
 luminosity & $3.0\%$ & $ 1.5\%$ \\
 PDF uncertainty & $17\%$ & $8.3\%$ \\
 NLO corrections & $16\%$ & $8.0\%$ \\
 sparticle mass $\Delta \tilde{m}=10$ GeV & $11\%$ & $5.6\%$ \\
 statistics (optimistic) & $8.6\%$  & $4.3\%$ \\
 statistics (conservative) & $23\%$ & $11\%$ \\
 \hline
 total (optimistic) & $27\%$ & $14\%$ \\
 total (conservative) & $35\%$ & $17\%$ \\
 \hline
  \end{tabular}
\caption{\label{errors_mSUGRA} Relative errors for the signal monojet cross section
(second column) and the $\widetilde{\chi}_1^0 \widetilde{q}_R q$ coupling $\lambda$
(third column) from different sources (first column). The numbers are for the
mSUGRA benchmark scenario, eq.~(\ref{benchmark_mSUGRA}).}
}

In Table~\ref{errors_mSUGRA}, we show the statistical and the most important
systematic errors for a measurement of the signal cross section 
and $\widetilde{\chi}_1^0 \widetilde{q}_R q$
coupling $\lambda$.  The total error is obtained by combining
all errors in quadrature.
The errors were estimated as follows: 
\begin{itemize}
\item In order to determine the monojet cross section from the observed number of events, one
needs to know the total number of collisions, i.e.~the integrated luminosity. According
to Ref.~\cite{lumi:2008zzj,Efthymiopoulos:2005ik,Boonekamp:2004tk},
the integrated luminosity at LHC will be measured to a precision of $3\%$ or better.
\item To translate the hadronic cross section to a parton-level cross section,
one needs to know the parton distribution functions (PDFs) accurately. We estimated the
error from the PDFs by comparing the hadronic cross sections with
different PDF releases; namely CTEQ6l, CTEQ6ll, CTEQ6m \cite{Pumplin:2002vw}, MSTW2008lo, MSTW2008nlo,
MSTW2008nnlo \cite{Martin:2009iq}, GJR08 \cite{Gluck:2007ck,Gluck:2008gs} and
a cubic interpolation of MRST LO** \cite{Sherstnev:2007nd}
(default {\tt Herwig++} PDF). We found the largest variation between MRST LO**
and MSTW2008lo. It corresponds to a variation of $\pm 17\%$ around the central
value of the total hadronic signal cross section.
We expect the PDF errors to be reduced significantly after the input to PDFs of
LHC data, but for now we use the current value of the error.
\item Further uncertainties arise from the unknown next-to-leading
  order (NLO)
corrections, which are not yet fully implemented \cite{Plehn:2004rp}.
The corresponding uncertainty is estimated 
by varying the renormalization scale $Q$ between
$\tfrac{1}{4}(m_{\tilde{q}_R}+m_{\tilde{\chi}_1^0})<Q<m_{\tilde{q}_R}+m_{\tilde{\chi}_1^0}$
(where the average squark mass of the right-handed first and second generation squarks
is denoted by $m_{\tilde{q}_R}$)
leading to an error of $+16\%/-13\%$
for the total LO cross section. In Table~\ref{errors_mAMSB}, we use $16\%$ in
order to be rather conservative, with the knowledge that the NLO calculation
would be ready and significantly decrease (possibly by a factor of a few) this
uncertainty.
The LO cross section was derived with {\tt Prospino2.1}
\cite{Prospino}.
\item
The relatively light SUSY spectrum in Table~\ref{mSUGRA_scenario} leads to
copious production of SUSY
particles at the LHC \cite{Baer:2010tk}. In this case, we expect that the $\widetilde{\chi}_1^0$ and squark masses can be reconstructed
to a precision of at least $10$ GeV \cite{Weiglein:2004hn}.
Varying the squark and $\widetilde{\chi}_1^0$ masses by $\pm 10$ GeV introduces an error of $\pm 11\%$ around
the central value of the total monojet cross section.
\item We always get an error due to statistical fluctuations of the SM backgrounds. In Table~\ref{errors_mAMSB},
we present two statistical estimates for this error as described in Section~\ref{Sec:backgrounds}.
The optimistic (conservative) estimator is $\sqrt{B}$ ($\sqrt{7\,B}$), with $B$ the number of SM background events.
\end{itemize}

We have neglected in Table~\ref{errors_mSUGRA} errors from the SUSY background, because an precise estimate
of its systematic uncertainties lies beyond the scope of this publication. However, we expect them
to be small, because it is in principle possible to extrapolate them from data. As we observe in
Fig.~\ref{Fig:m_mSUGRA}, the invariant mass distribution of the monojet is dominated by the SUSY background
for $m({\rm jet})>70$ GeV providing a control sample for a fit of the SUSY background.
Further model assumptions will also reduce the systematic uncertainties.

We can see in Table~\ref{errors_mSUGRA} that the biggest uncertainties are: the
PDF uncertainty, NLO corrections and from the SM statistics (conservative
estimate). However, one can
hope that the uncertainty of the PDFs will rapidly decrease after LHC has
taken and
analyzed some first data. Furthermore, the calculations of higher order corrections might improve in the future.
Unfortunately, an integrated luminosity of $300\,{\rm fb^{-1}}$ corresponds already to the maximum expected
LHC data. Some improvement might be possible by combining the data of both multi purpose detectors.

To conclude, a measurement of $\lambda$ via monojet production at the LHC is in principle possible
for scenarios with a bino-like $\widetilde{\chi}_1^0$ LSP like mSUGRA. However, at least parts of the
SUSY mass spectrum need to be relatively light, i.e.~the right-handed up-squark and down-squark
($\widetilde{\chi}_1^0$ LSP) mass should be $\lesssim 500$~GeV ($\lesssim 100$~GeV).
A measurement of the $\widetilde{\chi}_1^0 \widetilde{q}_R q$ coupling at the
$10\%$-level
is feasible by the end of LHC running.  The bino couplings to quark-squark pairs is
proportional to the corresponding squark hypercharge $Y$, which is different
for ${\widetilde d}_R$, and ${\widetilde u}_R$.
Since production of both squarks contributes to our SUSY monojet signal,
this fact must be taken into account when extracting a value of $\lambda$ from the 
data.\footnote{We expect that the first generation squarks are roughly mass-degenerate.
However, if there were a significant hierarchy between the 
up and down-type squark masses, then one would
dominantly produce the lightest squark, and the extraction of $\lambda$
from data would be more straightforward.}
For much larger masses, the monojet
cross section is too small to allow its extraction.


The situation is much more promising for SUSY scenarios
with a wino-like $\widetilde{\chi}_1^0$ LSP. The monojet
cross section is now enhanced compared to the bino-LSP case due to larger
gauge couplings and
due to additional diagrams involving a charged wino.

\section{Reconstruction of the $\widetilde{\chi}_1^0\widetilde{q}q$ Coupling for
a Wino LSP}
\label{Sec:wino_coupling}

\subsection{Benchmark Scenario}
\label{Sec:bechmark_mAMSB}

The second benchmark scenario is a mAMSB scenario \cite{anomalymed,AMSB,Gherghetta:1999sw} with a wino-like $\widetilde{\chi}_1^0$
LSP and with a wino-like $\widetilde{\chi}_1^+$ NLSP that is nearly degenerate in mass
with the $\widetilde{\chi}_1^0$ LSP. The scenario is described by the parameters
\begin{eqnarray}
M_{3/2}=33\,{\rm TeV}, \,\, M_0=200 \, {\rm GeV}, \,\, \tan \beta=10, \,\, {\rm sgn}(\mu)=+1 \,.
\label{mAMSB_param}
\end{eqnarray}
The resulting sparticle masses at the electroweak scale are given in Table~\ref{mAMSB_scenario}.
\TABLE{
\begin{tabular}{|cc|cc|}
 \hline
 sparticle & mass [GeV] & sparticle & mass [GeV] \\
 \hline
 $\widetilde{\chi}_1^0$ & 106.5 &  $\widetilde{\chi}_4^0$ & 593 \\
 $\widetilde{\chi}_1^+$ & 106.7 &  $\widetilde{\chi}_2^+$ & 594 \\
 $\widetilde{\tau}_1$ & 113 &  $\widetilde{b}_1$ & 634 \\
 $\widetilde{\nu}_\tau$ & 135 & $\widetilde{t}_2$ & 688 \\
 $\widetilde{\nu}_e/\widetilde{\nu}_\mu$ & 138 & $\widetilde{u}_L/\widetilde{c}_L$ & 722 \\
 $\widetilde{e}_R/\widetilde{\mu}_R$ & 150 & $\widetilde{b}_2$ & 723\\
 $\widetilde{e}_L/\widetilde{\mu}_L$ & 159 &  $\widetilde{d}_L/\widetilde{s}_L$ & 726\\
 $\widetilde{\tau}_2$ & 179 &    $\widetilde{u}_R/\widetilde{c}_R$ & 726 \\
 $\widetilde{\chi}_2^0$ & 298 &   $\widetilde{d}_R/\widetilde{s}_R$ & 732 \\
 $\widetilde{t}_1$ & 521 &   $\widetilde{g}$ & 745\\
 $\widetilde{\chi}_3^0$ & 584 &  &  \\
 \hline
 \end{tabular}
\caption{\label{mAMSB_scenario}
Supersymmetric particle masses for the mAMSB scenario defined in eq.~(\ref{mAMSB_param}).
We show the masses of the wino-like $\widetilde{\chi}_1^0$ and $\widetilde{\chi}_1^+$ up to a precision of four digits,
because they are nearly degenerate in mass. The sparticles are ordered by
mass.}
}

In this scenario, the total signal cross-section is $\sigma(pp \rightarrow
{\widetilde q}_L {\widetilde \chi}_1^0/\widetilde{\chi}_1^+)=470$~fb, and
$\lambda=0.99 g$ due to the small bino/higgsino
admixture. The lightest chargino, which will be included in our signal, couples
with strength $0.98g$ to ${\widetilde d}_Lu$ because of a small higgsino admixture and
with $1.00 g$ to ${\widetilde u}_Ld$. 
The small differences between the ${\widetilde \chi}_1^\pm$ and ${\widetilde \chi}_1^0$
couplings are much smaller than our measurement errors, and shall be neglected
in the following.

\subsection{Event Numbers and Cuts}
\label{Sec:cuts_mAMSB}

In this section we apply the same procedure as we did in Section~\ref{Sec:cuts_mSUGRA}.
We develop a set of cuts such that a clear signal over the SM backgrounds is visible.
Furthermore, we want to obtain a good signal to SUSY background ratio, which is needed
for a precise estimation of the $\widetilde{\chi}_1^0 \widetilde{q}_L q$ coupling $\lambda$.
There are some differences in our analysis to the bino-LSP scenario, eq.~(\ref{benchmark_mSUGRA}).
Instead of an integrated luminosity of 300 ${\rm fb}^{-1}$, we assume only 100 ${\rm fb}^{-1}$ to show
that less statistics is needed to reconstruct $\lambda$. This is due to the much larger
monojet cross section for wino-LSP scenarios; cf.~Section~\ref{signal_process}.
For the same reason the discovery potential is better, i.e.~we can investigate
a heavier spectrum; see Table~\ref{mAMSB_scenario} and
Table~\ref{mSUGRA_scenario}.

The cut flow for the mAMSB benchmark scenario,
eq.~(\ref{mAMSB_param}), is given in Table~\ref{cutflow_mAMSB}. The
first two cuts are identical to those in Table~\ref{cutflow_mSUGRA}.
However, the jet veto is slightly different (third row in
Table~\ref{cutflow_mAMSB}). Instead of vetoing additional jets with
$p_T$s larger than 30 GeV, we now relax this cut to 50 GeV.

\TABLE{
\begin{tabular}{|l|ccc|c|}
 \hline
 cut & all SM &  SUSY bkg. & signal & $S/\sqrt{B}$ \\
 \hline
 trigger & $3.81 \times 10^{7}$ & $1.04 \times 10^6$ & $44\,100$ & - \\
 lepton veto & $2.52 \times 10^7$ & $621\,000$ & $43\,800$ & - \\
 $p_T$(jet2) $<$ 50 GeV & $1.73 \times 10^7$ & $111\,000$ & $16\,200$ & 3.9 (1.5) \\
 $\ptmiss$ $>$ 300 GeV & $171\,000$ & $11\,000$ & $8\,390$ & 20 (7.7) \\
 $m$(jet1) $<$ 80 GeV & $135\,000$ & $6\,020$ & $6\,370$ & 17 (6.5) \\
 \hline
 tau veto & $119\,000$ & $5\,840$ & $6\,370$ & 18 (7.0) \\
 $b$-jet veto & $115\,000$ & $5\,290$ & $6\,320$ & 19 (7.0) \\
 \hline
  \end{tabular}
\caption{\label{cutflow_mAMSB} Same as Table~\ref{cutflow_mSUGRA}, but now for the mAMSB
benchmark scenario, eq.~(\ref{mAMSB_param}). We have assumed an integrated luminosity
of only 100 ${\rm fb}^{-1}$ at $\sqrt{s}=14$ TeV.  Jet1 and jet2 denote the jet with
the largest $p_T$ and second-largest $p_T$, respectively.}
}

\FIGURE[t]{
\unitlength=1cm
\begin{picture}(9,8.5)
\put(-1.3,0.34){
\epsfig{file=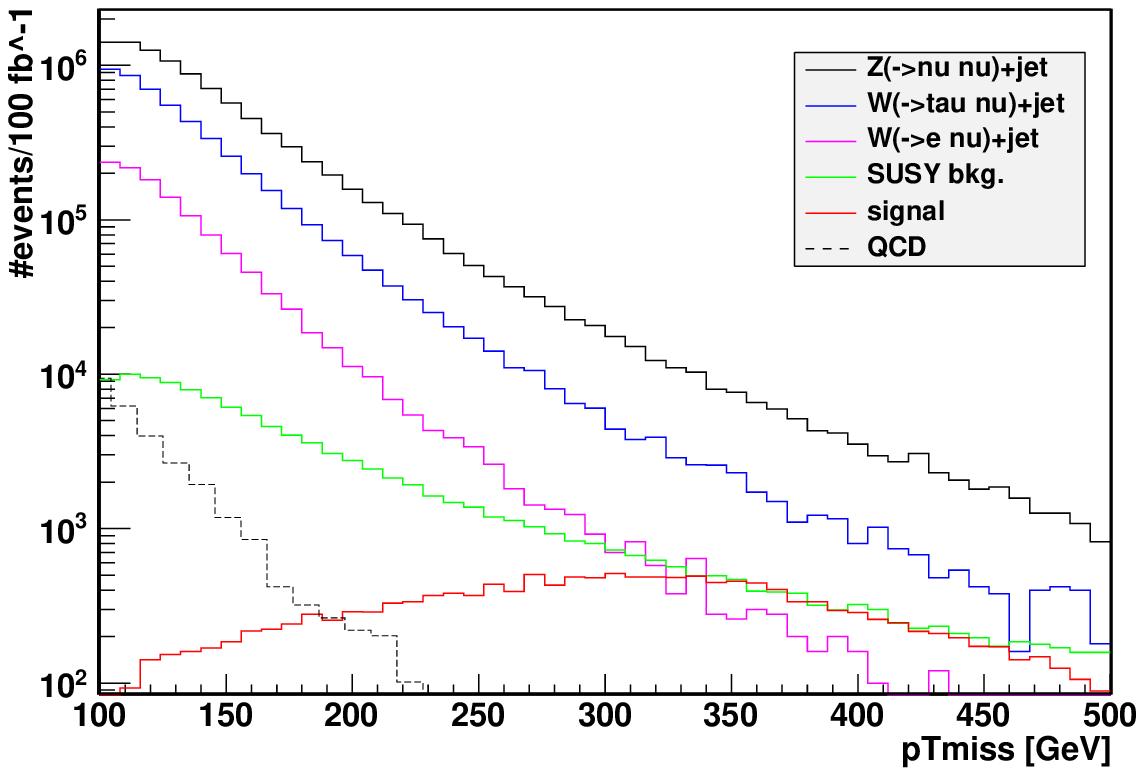,width=11.5cm}
}
\end{picture}
\caption{Missing transverse momentum distribution for the $Z(\rightarrow \nu \bar \nu)$+jet
(black histogram), $W(\rightarrow \tau \nu)$+jet (blue histogram), $W(\rightarrow e \nu)$+jet
(magenta histogram), the SUSY background (green histogram), the QCD background
  (black dashed histogram) and the signal process (red histogram).
The first three cuts of Table~\ref{cutflow_mAMSB} have been applied. The
number of events correspond
to an integrated luminosity of 100 ${\rm fb}^{-1}$ at $\sqrt{s}=14$ TeV. We
assume the mAMSB scenario of eq.~(\ref{mAMSB_param}).}
\label{Fig:pts_mAMSB}
}

In contrast to the mSUGRA scenario, the SUSY background for our mAMSB benchmark point
is dominated by Drell-Yan wino pair production; cf.~Section~\ref{Sec:backgrounds}.
At the parton level, the event topology is similar to that of the signal and to
that of the dominant SM
background: two invisible particles recoiling against a single quark or gluon. After
initial and final state radiation, additional jets can arise and the event
might fail the jet
veto. The veto then reduces the relative number of signal and
($Z$+jet and wino-pair plus jet) background events equally due to the similar event topologies. Therefore, the jet veto reduces the significance $S/\sqrt{B}$ whereas the
signal to SUSY
background ratio stays roughly constant. A weaker jet veto thus increases the
visibility of the signal. Note that for the mSUGRA scenario, eq.~(\ref{benchmark_mSUGRA}),
the topology of most of the SUSY background is different. For example, in the
mSUGRA scenario, the dominant background is
squark pair production,
where both squarks decay into a quark and a $\widetilde{\chi}_1^0$.

\FIGURE[t]{\\
\epsfig{file=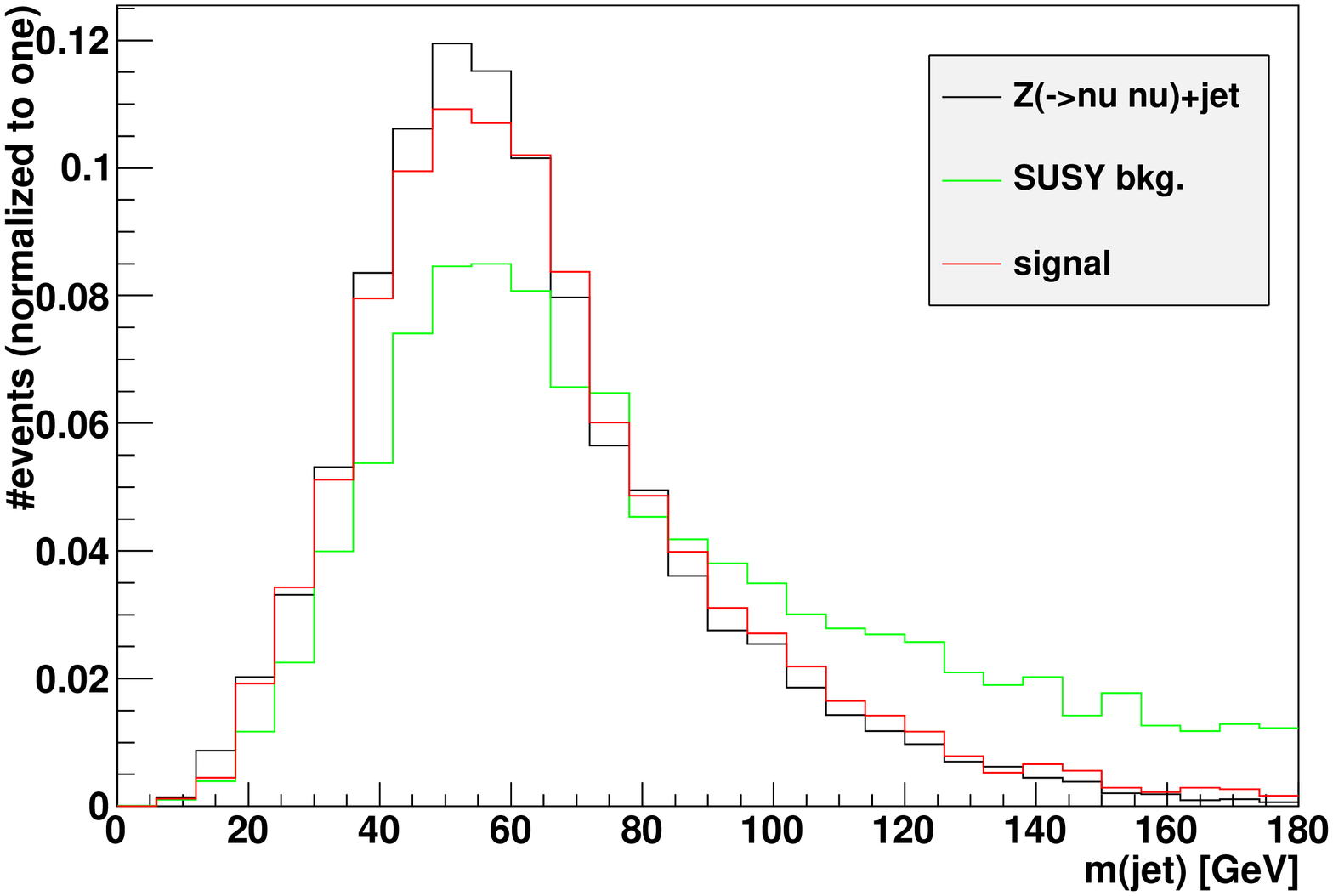,width=7.7cm,angle=270}
\caption{Invariant mass distribution of the hardest jet for the
the $Z$+jet background (black histogram), the SUSY background (green histogram) and
the signal process (red histogram). The first four cuts of Table~\ref{cutflow_mAMSB}
have been applied. The distributions are normalized to one. For the signal, we
assume the mAMSB scenario
of eq.~(\ref{mAMSB_param}). The QCD background has a distribution almost
indistinguishable
from that of $Z(\rightarrow \nu \bar \nu)$+jet.
}
\label{Fig:m_mAMSB}
}

After the jet veto, we still observe in Table~\ref{cutflow_mAMSB} an overwhelming SUSY background,
i.e.~the signal to SUSY background ratio is roughly seven whereas for our
mSUGRA benchmark point this ratio was less than two; see Table~\ref{cutflow_mSUGRA}.
However a great improvement is possible with a cut on the $\ptmiss$.
The relevant distributions (after the first three cuts in Table~\ref{cutflow_mAMSB} were applied)
are shown in Fig.~\ref{Fig:pts_mAMSB}. Here, the SUSY background distributions as well as the
SM background distributions fall off exponentially. In contrast, the $\ptmiss$ distribution of
the SUSY background in Fig.~\ref{Fig:pts_mSUGRA} does not show the same
exponential fall-off; a clear sign that the SUSY background is different in
the mSUGRA and mAMSB cases. Applying again a lower $\ptmiss$ cut around the
Jacobian peak of the signal (the red histogram in Fig.~\ref{Fig:pts_mAMSB})
increases the signal
to SUSY background ratio to 0.8 and the optimistic (conservative) significance
to 20 (7.7). Note that we have also applied the same cut on the $p_T$ of the
hardest jet. We also see that for this point, the required level of
cancellation of the SM model backgrounds is at the $\sim 10\%$ (or
larger) level.
Deriving a systematic error smaller than this on the backgrounds measurement
ought to be easily achievable with the high number of data expected.

In Fig.~\ref{Fig:m_mAMSB}, we present the invariant mass distribution of the
hardest jet
after the first four cuts of Table~\ref{cutflow_mAMSB} were applied. Because the SUSY background
is dominated by Drell-Yan like wino pair production, the $m({\rm jet1})$ distribution is less
flat than those in Fig.~\ref{Fig:m_mSUGRA}. However, an improvement with an upper cut on $m({\rm jet1})$
is still possible. Demanding $m({\rm jet1})<80$ GeV increases the signal to SUSY background ratio
to 1.1. At the same time, the signal significance is not significantly changed. It is 17 (6.5)
for our optimistic (conservative) estimate.
We finally show in the last two columns of Table~\ref{cutflow_mAMSB} the effect of a tau and $b$-jet veto.
Again, we observe no great improvement by such a veto and thus we shall not apply it in the
following.

\subsection{Coupling Reconstruction}
\label{Sec:coupling_reco_mAMSB}

We now show to which precision the $\widetilde{\chi}_1^0 \widetilde{q}_L q$ coupling $\lambda$ of the mAMSB benchmark scenario
of Section~\ref{Sec:bechmark_mAMSB} can be reconstructed with an integrated luminosity of
$100\, {\rm fb}^{-1}$ at $\sqrt{s}=14$ TeV. We employ the cuts given in Section~\ref{Sec:cuts_mAMSB}
without the tau and $b$-jet veto. We will closely follow the procedure employed in
Section~\ref{Sec:coupling_reco_mSUGRA}. We reconstruct $\lambda$ under the
assumption of the SUSY backgrounds of our mAMSB point.
We exhibit in Table~\ref{errors_mAMSB} the statistical and the most important systematic errors for a measurement
of the cross section and thus the $\widetilde{\chi}_1^0 \widetilde{q}_L q$ coupling.

\TABLE[t]{
\begin{tabular}{|l|cc|}
 \hline
 error & $\Delta \sigma_{\rm mono}/\sigma_{\rm mono}$ &
$\Delta \lambda/\lambda$\\
 \hline
 luminosity & $3\%$ & $ 1.5\%$ \\
 PDF uncertainty & $17\%$ & $8.3\%$ \\
 NLO corrections & $18\%$ & $9\%$ \\
 sparticle mass $\Delta \tilde{m}=10$ GeV & $7.3\%$ & $3.7\%$ \\
 statistics (optimistic) & $5.8\%$  & $2.9\%$ \\
 statistics (conservative) & $15\%$ & $7.7\%$ \\
 \hline
 total (optimistic) & $26\%$ & $13\%$ \\
 total (conservative) & $30\%$ & $15\%$ \\
 \hline
  \end{tabular}
\caption{\label{errors_mAMSB} Relative errors for the signal monojet cross section (second column)
and the $\widetilde{\chi}_1^0 \widetilde{q}_L q$ coupling (third column) from different sources (first column).
The numbers are for the mAMSB benchmark scenario, eq.~(\ref{mAMSB_param}).}
}

Note that errors from the SUSY background are not included.
For our mAMSB benchmark scenario (and for many wino LSP scenarios in general),
the SUSY background is dominated by wino pair production plus jet. The differential
distribution as a function of the missing transverse momentum is known \cite{Giudice:2010wb}.
One can also measure the SUSY background below the Jacobian peak of the monojet distribution.
For our mAMSB benchmark scenario, we found that the SUSY background in an $\ptmiss$ interval of 150 GeV
to 300 GeV is more than six times larger than the monojet signal as can be seen in Fig.~\ref{Fig:pts_mAMSB}.
At the same time, for the total number of SUSY events we found a (conservative) significance
over SM background of roughly ten. Therefore,
it should be possible to estimate   the SUSY background from data which reduces
significantly the systematic uncertainties of the SUSY background. Assumptions about
the underlying model can further reduce this error. We thus expect
that errors from SUSY background will not significantly alter the results
in Table~\ref{errors_mAMSB}. However, a precise reconstruction of the SUSY
background lies beyond the scope of this publication.

As in Section~\ref{Sec:coupling_reco_mSUGRA} the largest errors come from the PDF uncertainty, from the NLO corrections
and from the SM statistics (conservative estimate). There is a good chance that the PDF uncertainty will decrease after LHC has
analyzed some first data. Also the calculations of some of the unknown higher
order corrections might be performed
in the future. Finally, increasing the integrated luminosity would decrease
the statistical error.
In contrast to the bino LSP scenario, where we already assumed an integrated
luminosity
of 300 fb$^{-1}$, the coupling reconstruction is possible earlier for our wino LSP scenario, because
we only assumed 100 fb$^{-1}$ of data.
For example, assuming an integrated luminosity of $300\,{\rm fb^{-1}}$ and an error
from the PDFs and from higher order corrections of $10\%$, the relative error of $\lambda$
would decrease from $13\%$ ($15\%$) in Table~\ref{errors_mAMSB} to $8.3\%$ ($9.2\%$) for
the optimistic (conservative) statistical estimate. To conclude, a measurement of $\lambda$ with a precision
of roughly $10\%$ at the end of LHC running seems to be possible for our mAMSB benchmark scenario.

\subsection{Reconstruction of Couplings under the Assumption of a Wino LSP}
\label{Sec:parameter_scan}

We have seen that the reconstruction of the coupling $\lambda$ is much more
promising
in SUSY scenarios with a wino like LSP than with a bino like LSP. Therefore, we will
henceforth concentrate
in this section on wino LSP scenarios. We will investigate the precision
to which $\lambda$ can be measured as a function of the (left-handed) squark mass
and the $\widetilde{\chi}_1^0$ mass.

For that purpose, we perform a two-dimensional parameter scan. We have
varied the pole masses of the left-handed first generation squarks (winos)
between 400 GeV and 1210 GeV (100 GeV and 400 GeV). All other SUSY particles
are decoupled. This part of parameter space has the chargino-squark-quark
couplings equal to $\lambda=g$ to better than per mille precision. 
We employ a grid of ten times ten points equally spaced in the
squark--wino mass plane. 
We assumed an integrated  luminosity of 100 ${\rm fb}^{-1}$
at $\sqrt{s}=14$~TeV. For the relative error from the
luminosity we use again $3\%$, whereas the errors from PDFs, NLO corrections,
and the SUSY mass uncertainties were estimated in the following way.
We calculated for a sub-grid with four times four parameter points the
respective 
errors as it was done in Section~\ref{Sec:coupling_reco_mAMSB} and extrapolated
the errors for the other parameter points.

The PDF uncertainties are described well by a linear function of the squark and
$ \widetilde{\chi}_1^0$ mass sum in which the error {\it decreases} with increasing mass.
This behaviour can be understood by noting that the error of the gluon PDF decreases
in the relevant region for increasing Bjorken-$x$ \cite{Martin:2009iq}.
For the NLO uncertainties we use a logarithmic fit as a function of the sum
of the squark and $\widetilde{\chi}_1^0$ mass. {\tt Prospino} employs the squark
and $\widetilde{\chi}_1^0$ mass sum divided by two as its default factorization and renormalization scale.
For the scale variation both scales are then varied between 1/2 and 2 times the default value,
cf.~Section~\ref{Sec:coupling_reco_mSUGRA}. A logarithmic behaviour is expected,
because the scale dependence enters via loops.

The estimate of the error from SUSY mass uncertainties is more involved. In general,
the precision to which the masses can be reconstructed depends on the nature of the
SUSY particle, the mass spectrum and the model assumptions; see for example
Refs.~\cite{Weiglein:2004hn,Bechtle:2009ty}. We here employ a simple approach.
We assume that a squark mass of $m_{\tilde{q}_L}=720$ GeV can be reconstructed with an absolute error
of $\delta m_{\tilde{q}_L}=10$ GeV. For a $\widetilde{\chi}_1^0$ with a mass of 100 GeV we assume the same error.
This corresponds to our assumption made in Section~\ref{Sec:coupling_reco_mAMSB}.
We then estimated the error for other squark masses $m'_{\tilde{q}_L}$ via
\begin{eqnarray}
\delta m'_{\tilde{q}_L} = \delta m_{\tilde{q}_L} \times \frac{m'_{\tilde{q}_L}}{m_{\tilde{q}_L}} \times
\sqrt{\frac{\sigma(PP\rightarrow \widetilde{q}_L \widetilde{q}_L)}{\sigma(PP\rightarrow \widetilde{q}'_L \widetilde{q}'_L)}}
\, ,
\label{mass_error_estimate}
\end{eqnarray}
i.e.~the relative error scales with the square root of the inverse squark pair
production cross section, $\sigma(PP\rightarrow \widetilde{q}'_L \widetilde{q}'_L)$,
as expected if the errors are dominated by statistical uncertainties
and if the masses are reconstructed from cascade decays of squarks.

\FIGURE[t]{
\epsfig{file=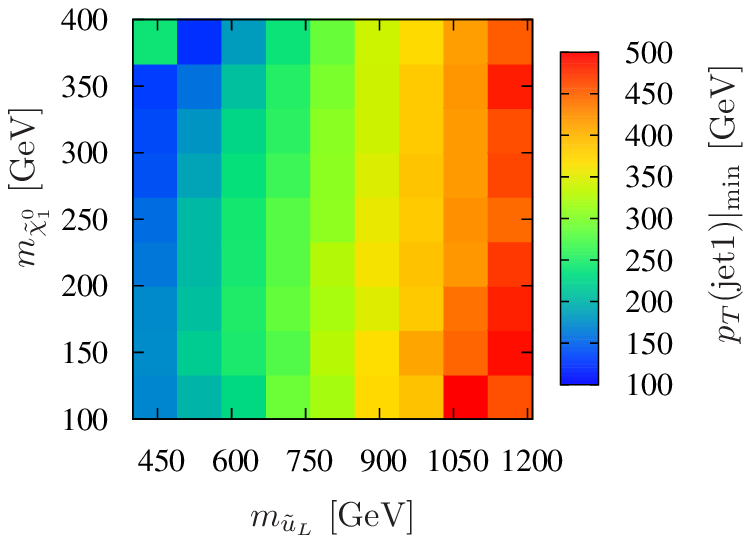,width=7.4cm}
\epsfig{file=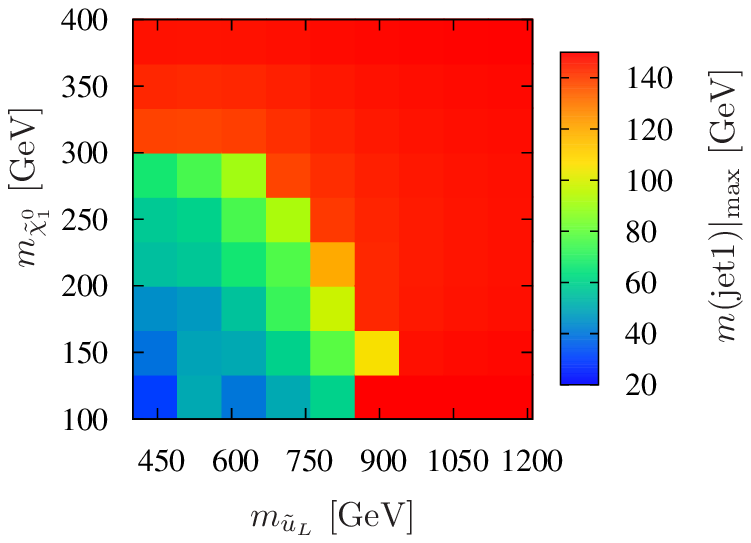,width=7.4cm}
\caption{\label{Fig:cuts} Lower cut for the jet-$p_T$ (left-hand side) and upper cut for
the jet invariant mass (right-hand side) as a function of the squark and
$ \widetilde{\chi}_1^0$ mass. The red region in the right figure corresponds to
no cut on $m({\rm jet1})$.}
}

We also employ the same absolute error for the $\widetilde{\chi}_1^0$
mass, because the squark and $\widetilde{\chi}_1^0$
mass errors are always strongly correlated to each other if the squark
decays into the $\widetilde{\chi}_1^0$ plus a jet
\cite{Bechtle:2009ty}.  Moreover, analyses such as the
benchmark point SPS1a~\cite{AguilarSaavedra:2005pw} show that the
lightest neutralino mass can be slightly better reconstructed than the
squark masses \cite{Weiglein:2004hn,Bechtle:2009ty}. 
We therefore follow a conservative approach in this analysis.  
{}From the mass errors, we obtain errors
for the total signal cross section following the procedure described in
Section~\ref{Sec:coupling_reco_mAMSB}. We found that a quadratic fit
as a function of the squark mass describes very well the calculated
errors as might be expected from eq.~(\ref{mass_error_estimate}).

In order to find the right cuts for each SUSY scenario, we have employed a similar strategy
to the one in Section~\ref{Sec:coupling_reco_mAMSB}. We applied the first
three cuts of Table~\ref{cutflow_mAMSB}. Starting from the trigger cut
of 100 GeV, we searched numerically
in steps of 10 GeV for a (lower) $p_T$ cut on the hardest jet that
maximizes the significance $S/\sqrt{B}$, where
$S$ ($B$) corresponds to the number of signal (SM background events).
Finally, we varied the (upper) cut on the invariant mass of the hardest jet in steps of 10 GeV
in order to maximize the signal to SUSY background ratio. However, we only employ
a cut on the invariant mass as long as the (conservative) significance
is larger than 
five.

The resulting cuts employed are shown in Fig.~\ref{Fig:cuts} for the jet-$p_T$ (which is equal to the
cut on $\ptmiss$) and for the jet invariant mass. They are given as a function
of the left-handed  up-squark mass, $m_{\tilde{u}_L}$, and the (wino-like) $\widetilde{\chi}_1^0$ LSP mass,
$m_{\tilde{\chi}_1^0}$. Note that the left-handed down squark
and left-handed up squark are quasi-degenerate, as are the
$ \widetilde{\chi}_1^+$ and $\widetilde{\chi}_1^0$. We can see that the lower cut on the jet-$p_T$ increases with
increasing squark mass and decreasing $\widetilde{\chi}_1^0$ mass as expected from the position
of the Jacobian peak given in \eq{Jpeak}.
Recall that we usually obtain the best significance
if we apply the jet-$p_T$ cut close to the Jacobian peak; cf.~Sections~\ref{Sec:cuts_mSUGRA}
and \ref{Sec:cuts_mAMSB}.

However, an exception is the quite large jet-$p_T$ cut in the upper left corner of
Fig.~\ref{Fig:cuts}. Due to the small squark--$\widetilde{\chi}_1^0$ mass
difference, the Jacobian peak lies below $\lesssim 100$ GeV, {\it i.e} below the trigger
cut. For larger $p_T$ values (compared to the peak), the jet-$p_T$ (and $\ptmiss$) distribution of the signal
falls off less steep than the SM backgrounds. Therefore, for a small squark--$\widetilde{\chi}_1^0$ mass difference
the best cut will lie above the Jacobian peak.

Fig.~\ref{Fig:cuts} (right side) shows the upper cut on the invariant mass
of the monojet. The red region corresponds to parameter points, where
the conservative significance, $S/\sqrt{7B}$, is smaller than five. In this case,
we do not apply a cut on the invariant jet mass in order not to suppress
the signal cross section.
We observe that harder cuts on the invariant mass are usually applied for smaller squark
and $\widetilde{\chi}_1^0$ masses, because the number of signal events is large and
a hard cut will not suppress the signal below the discovery reach. Recall from
Fig.~\ref{Fig:m_mSUGRA} and Fig.~\ref{Fig:m_mAMSB} that harder cuts on the invariant
mass lead often to a better signal to SUSY background ratio.

\FIGURE[t]{
\epsfig{file=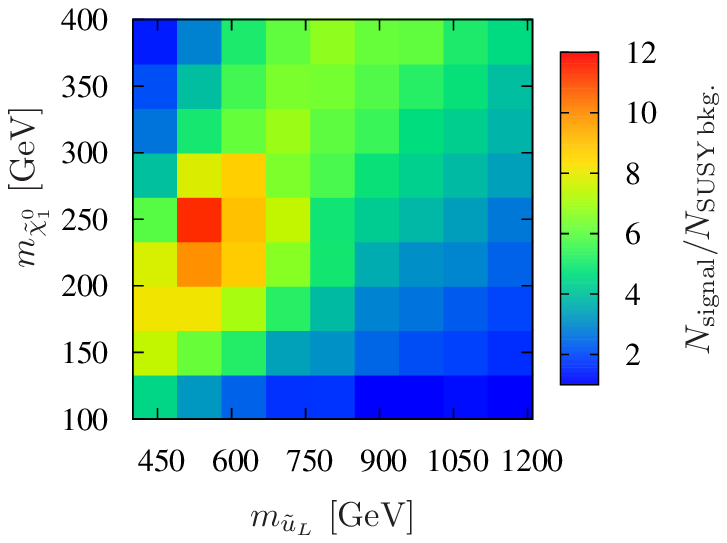,width=3.4in}
\caption{\label{Fig:signal_vs_SUSYbkg} Ratio of the number of signal
events to the number of SUSY background events as a function of the
squark and $\widetilde{\chi}_1^0$ mass.}
}

The importance of the invariant mass cut can be seen in Fig.~\ref{Fig:signal_vs_SUSYbkg},
where the ratio of signal to SUSY background events (after cuts) as a function of the
squark and $\widetilde{\chi}_1^0$ mass is shown. We observe that the ratio varies between 1.2 and 12.
This good signal to SUSY background ratio justifies our approximation to neglect the
(unknown) error from SUSY backgrounds as long as the SUSY background can be determined
to a precision of a few $10\%$; see also the discussion in Section~\ref{Sec:coupling_reco_mSUGRA}
and Section~\ref{Sec:coupling_reco_mAMSB}.

We also see some interesting structure in Fig.~\ref{Fig:signal_vs_SUSYbkg}: For fixed
squark mass, the signal to SUSY background ratio first increases when we increase the
$ \widetilde{\chi}_1^0$ mass and then decreases again. For small $\widetilde{\chi}_1^0$ mass,
the SUSY background is dominated by Drell-Yan wino pair production. However, when the
$ \widetilde{\chi}_1^0$ mass is increased the respective cross section will decrease and
squark pair production will take over to be the dominant background. At the same time,
as long as the $\widetilde{\chi}_1^0$ mass increases moderately (relative to the squark mass)
the signal cross section decreases less fast than the SUSY background leading to a better
signal to SUSY background ratio. However, when we come to the region where
$m_{\tilde{\chi}_1^0} \approx m_{\tilde{q}_L}$ we obtain a reduced signal to SUSY background ratio.
%
The signal to SUSY background ratio can be further increased with a
cut on the jet invariant mass. As can be seen in Fig.~\ref{Fig:signal_vs_SUSYbkg}, this
can result in a signal to SUSY background ratio larger than ten!

\FIGURE[t]{
\epsfig{file=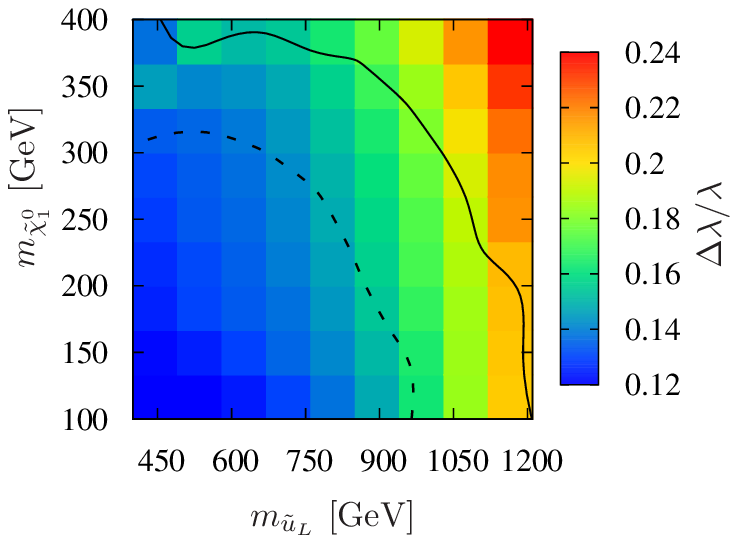,width=7.4cm}
\epsfig{file=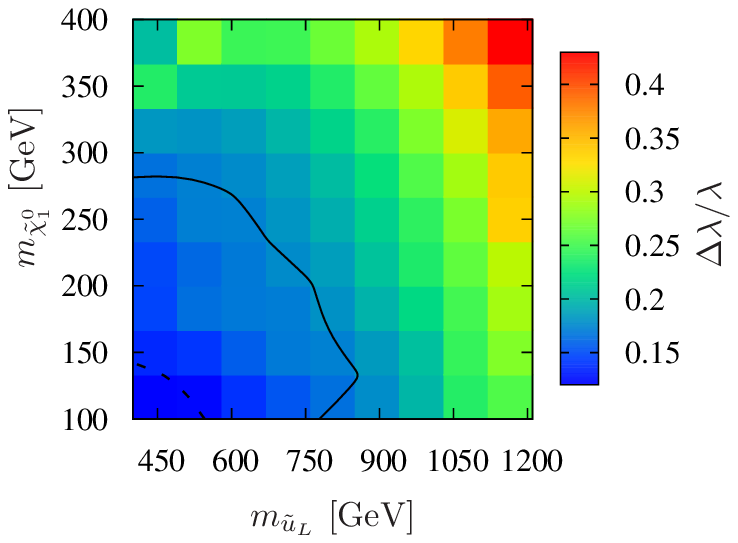,width=7.4cm}
\caption{\label{Fig:error_coupling} Fractional precision to which
the ${\widetilde\chi}_1^0 \widetilde{q}_L q$ coupling $\lambda$
can be reconstructed as function of the squark and
$ \widetilde{\chi}_1^0$ mass. The left (right) figure employs
our optimistic (conservative) estimate for the SM background
uncertainties. The solid and dashed black lines correspond to
$S/\sqrt{B}$ ($S/\sqrt{7B}$) of $5\sigma$ and $10\sigma$, respectively.}
}

The fractional precision that can be achieved
can be seen in Fig.~\ref{Fig:error_coupling},
where we show the relative error of the coupling $\lambda$ as a function of
the left-handed
up-squark mass and the $\widetilde{\chi}_1^0$ LSP mass.
Fractional precisions between 0.12 and 0.24 are possible throughout
parameter space for optimistic assumptions, or 0.12--0.44 for conservative ones.
The most precise measurements are for lighter sparticles,
where signal event numbers are higher.

The significance of the SUSY monojet signal is also given
in Fig.~\ref{Fig:error_coupling} by the solid (dashed) black line
corresponding to the $5\sigma$ ($10\sigma$) region. We see a strong dependence on
whether we assume optimistic or conservative error estimates. Whether or not
the signal can be seen over statistical fluctuations of the background to
the 5$\sigma$ level, the coupling $\lambda$ may be bounded.

To conclude, the LHC will be able to test the $\widetilde{\chi}_1^0\widetilde{q}_Lq$
coupling relation to a precision of $\mathcal{O}(10\%)$ in large regions of
the wino LSP parameter space as long as the squark masses are $\lesssim 1$ TeV.

\section{Summary and Conclusion}
\label{Sec:summary}

If signatures of new physics beyond the SM are discovered that are consistent with SUSY, a
program of measurements of the interactions of new states would be useful to
hypothesis test the SUSY interpretation and cross-correlate with other
measurements.
As a first step, we have investigated the feasibility of 
SUSY monojet production 
at the LHC via $q g \rightarrow {\widetilde q} {\widetilde \chi}_1^0 \rightarrow q
{ \widetilde \chi}_1^0 {\widetilde \chi}_1^0$ for measuring the 
${\widetilde{\chi}_1^0\widetilde q}q$ coupling $\lambda$, under the
assumption that
the ${\widetilde \chi}_1^0$ is the LSP. 

For a bino LSP, supersymmetry predicts 
that $\lambda$ is proportional to $g'$,
the U$(1)_Y$ gauge coupling.  
For a wino LSP however, supersymmetry predicts
$\lambda = g$.   
When added to other empirical
information on the field content of the neutralino, the value of the coupling
constitutes a test of supersymmetry.
In general, the MSSM ${\widetilde \chi}_1^0$ contains admixtures
of the wino,
the bino and the higgsino, resulting in $\lambda$ being a mixture of $g$,
$g'$ and $0$, respectively. 
Without additional information on the field content coming from
other measurements, a measurement
of the coupling of the LSP to a quark and a squark
(in the framework of the MSSM)
yields valuable information regarding the neutralino field content. As such, it
could be used in the
calculation of its thermal relic density in the universe, or for dark matter
direct detection rates, for example. 
Our signal would only be measured
significantly above backgrounds after data
have been amassed for many years at the LHC. Therefore it is reasonable to
expect that information on the mass spectrum would already be known, e.g. squark
and neutralino mass scales, thereby constraining the SUSY backgrounds.

At this stage, it would be useful to extract, as far as possible, SUSY
backgrounds from other data samples.  However, without data to guide 
us, the number of possibilities is very large.
The most difficult part of
estimating SUSY backgrounds would likely be
the extraction of the relevant branching ratio for the $\widetilde q \rightarrow q
{ \widetilde \chi}_1^0$ decay. 
Nevertheless, placing an upper bound of 1 on the branching ratio for an
observed SUSY monojet signal would yield a lower bound on
$\lambda$. If the LSP is a 
generic admixture of bino-wino-higgsino eigenstates, one must
also take into account the fact that the lightest chargino may couple
differently to the lightest neutralino. However, one would then generically
expect the ${\widetilde \chi}_1^\pm$ to not be quasi-mass degenerate with the
${\widetilde \chi}_1^0$, and
so its decays ought to be visible and therefore reducible by our cuts. 
On the other hand,
it may be that a simple model of supersymmetry breaking with only
a few parameters is
selected by fitting all available data. In this case, the branching ratio and
therefore the total rate of our monojet signal would
be predicted and would constitute an additional hypothesis
test on that model.

Measuring $\lambda$ is not easy at the LHC, but we have shown that
a fractional accuracy approaching $10\%$ is feasible with a
large data set,
providing a non-trivial and useful empirical constraint on the MSSM. As such,
it will be an important element of the proposed program of precision
measurements of new particle interactions to support the SUSY hypothesis.

\acknowledgments
B.C.A. would like to thank Mihoko Nojiri for early discussions and her
kind help with event generation, other
members of the Cambridge SUSY working group, Alan Barr, Jon Butterworth and
Kyle Cranmer for comments and
suggestions. H.E.H. appreciates useful conversations with Tilman Plehn, Herbi
Dreiner and JoAnne Hewett.
S.G. thanks Michael Kraemer, Jong Soo Kim, Sebastian Fleischmann and Peter
Wienemann for helpful discussions.  We are especially grateful to Laura Daniel
for pointing out an error in an earlier version of the Appendix.

B.C.A. is supported in part by a grant from STFC and an associateship from the
IPPP\@ in Durham.
The work of S.G. is supported
in part by the U.S. Department of Energy, under grant
number DE-FG02-04ER41268 and in part by a Feodor Lynen Research Fellowship
sponsored by the Alexander von Humboldt Foundation.  H.E.H. is supported
in part by the U.S. Department of Energy, under grant
number DE-FG02-04ER41268  and in part by a Humboldt Research Award sponsored by
the Alexander von Humboldt Foundation.

\appendix

\section{The Jacobian Peak in the Transverse Momentum Distribution}
\label{Sec:Jacobian}

The Jacobian peak is a well-known feature of the transverse momentum
distribution of the electron in the process $A+B\to W^\pm+X\to
e^\pm+\nu+X$, where $A$ and $B$ are the initial state hadrons.
The resulting peak at $p_T\simeq \frac{1}{2}m_W$ is a consequence of
the Jacobian that arises from changing kinematic variables
from $\cos\theta$ (where $\theta$ is the center-of-mass scattering
angle) to $p_T$.\footnote{For a pedagogical treatment, see Ref.~\cite{barger}.}
In this paper, we have focused on monojets that arise from $\widetilde q
{ \widetilde \chi}_1^0$ production, where $\widetilde q\to
q {\widetilde \chi}_1^0$, and the quark is observed as a hadronic jet.
The $p_T$ distribution of the quark jet also exhibits a Jacobian
peak. In this appendix, we derive an approximate expression for the
location of the peak in the transverse momentum distribution of the
jet.

Consider the $2\to 3$ scattering process, which schematically is of
the form:
\beq
a+b\to c+3\,,\,\text{followed by}~~c\to 1+2\,,
\eeq
where the decaying particle $c$ is spinless.  
Since the particles $a$, $b$ and $1$ represent light quarks or gluons,
we shall set their masses to zero, $m_a=m_b=m_1=0$.  We denote the
mass of particle $c$ (identified as the $\widetilde q$) 
to be $m_c\equiv M$, and the masses of
particles 2 and 3 (which are identified with $\widetilde\chi_1^0$)
to be $m_2=m_3\equiv m$.

If the particle $c$ is on-shell, then the corresponding
matrix element for the $2\to 2$ process, $a+b\to c+3$ is of the form
\beq \label{twotwo}
\mathcal{M}(a+b\to c+3)= C_1(s,t)\,,
\eeq
where $C_1(s,t)$ is a dimensionless function of
$s\equiv (p_a+p_b)^2$, $t\equiv (p_a-p_c)^2$ and the
particle masses.  The kinematical limits of $t$ are:
\beq
-\half\left[s-M^2-m^2+\lambda^{1/2}(s,M^2,m^2)\right]\leq t\leq
-\half\left[s-M^2-m^2-\lambda^{1/2}(s,M^2,m^2)\right]\,,
\eeq
where
\beq \label{lambdadef}
\lambda(a,b,c)\equiv a^2+b^2+c^2-2ab-2ac-2bc
\eeq
is the well-known triangle function of relativistic kinematics.
The squared matrix element for the decay of particle $c$ 
(which is either $\widetilde q_L$ or $\widetilde q_R$), 
summed over final spins, is given by
\beq \label{decay}
|\mathcal{M}(c\to 1+2)|^2= C_2 (M^2-m^2)\,,
\eeq
where $C_2$ is a 
dimensionless (real positive) constant that will eventually cancel
out in our computation.  Using, \eq{decay} it follows that
the total width of particle $c$ times the branching ratio is given by
\beq
B\Gamma= \frac{C_2 M}{16\pi}\left(1-\frac{m^2}{M^2}\right)^2\,,
\eeq
where the branching ratio $B\equiv B(c\to 1+2)$.

To set up our computation, we work in the center-of-mass system.  Then, the four-vectors of the
initial states and the observed final state (particle 1) are:
\beqa
p_a&=&\half\sqrt{s}(1\,;\,0\,,\,0\,,\,1)\,,\\
p_b&=&\half\sqrt{s}(1\,;\,0\,,\,0\,,\,-1)\,,\\
p_1&=&E_1(1\,;\,\sin\theta\,,\,0\,,\,\cos\theta)\,,
\eeqa
where $\theta$ is the scattering angle in the center-of-mass frame.
Following Ref.~\cite{byckling}, we define four Lorentz-invariant
quantities,
\beqa
t_1&\equiv & (p_a-p_1)^2=-\sqrt{s} E_1(1-\cos\theta)\,,\label{t1}\\
t_2&\equiv & t=(p_1+p_2-p_a)^2\,,\label{t2}\\
s_1&\equiv &p_c^2=(p_1+p_2)^2\,,\label{s1}\\
s_2&\equiv&(p_a+p_b-p_1)^2=s-2\sqrt{s} E_1\,.\label{s2}
\eeqa
We denote the three-body phase space integral by
\beq
R_3(s)\equiv\int\prod_{i=1}^3 \frac{d^3 p_i}{2E_i}\, \delta^{(3)}(\boldsymbol{p_a}+
\boldsymbol{p}_b-\boldsymbol{p}_1-\boldsymbol{p}_2-\boldsymbol{p}_3)\,
\delta(\sqrt{s}-E_1-E_2-E_3)\,.
\eeq
The key formula that we need is given by eq.~V-7.8 of
Ref.~\cite{byckling},
\beqa \label{basic}
\frac{dR_3}{ds_2 dt_1 ds_1}&=&\frac{\pi}
{8\lambda^{1/2}(s,m_a^2,m_b^2)\lambda^{1/2}(s,s_2,m_1^2)}
\,\Theta\{-G(s,t_1,s_2,m_a^2,m_b^2,m_1^2)\}\nonumber \\[6pt]
&&\hspace{1in}\times\Theta\{-G(s_1,s_2,s,m_2^2,m_1^2,m_3^2)\}
\int_0^{2\pi} d\phi\,,\label{lamdef}
\eeqa
where $G$ is the basic four-particle kinematic function first introduced
in Ref.~\cite{nyborg},
\beq
G(x,y,z,u,v,w)\equiv -\frac{1}{2}\,{\rm det}\begin{pmatrix}
2u & \quad x+u-v & \quad u+w-y \\
x+u-v & \quad 2x & \quad x-z+w \\
u+w-y & x-z+w & 2w\end{pmatrix}\,,
\eeq
and $\lambda$ is the triangle function defined in \eq{lambdadef}.
Expanding out the determinant yields the unwieldy
expression,\footnote{\Eq{G}, which was first defined
in Ref.~\cite{nyborg}, is also given in eq.~IV-5.23
of Ref.~\cite{byckling}.  We have noted a typographical error
in the latter; in the second line of eq.~IV-5.23, the first term $yzw$ should
read $yzv$.}
\beqa \label{G}
&& G(x,y,z,u,v,w)=xy(x+y)+zu(z+u)+vw(v+w)+x(zw+uv)+y(zv+uw) \nonumber \\
 &&\qquad\qquad\quad -xy(z+u+v+w)-zu(x+y+v+w)-vw(x+y+z+u)\,.
\eeqa
Finally, $\phi$ is the so-called \textit{helicity angle}~\cite{byckling}, 
which is most conveniently defined
in a reference frame where $\boldsymbol{\vec p}_2+\boldsymbol{\vec p}_3=
\boldsymbol{\vec p}_a+\boldsymbol{\vec p}_b-\boldsymbol{\vec p}_1=0$.
In this reference frame, $\phi$ is identified as
the azimuthal angle between the production plane spanned by
$\boldsymbol{\vec p}_b$ and $\boldsymbol{\vec p}_1$ and the
plane spanned by $\boldsymbol{\vec p}_1$ and $\boldsymbol{\vec p}_3$,
with $\boldsymbol{\vec p}_1$ as the axis.
The angle $\phi$ can be re-expressed in terms of the Lorentz-invariant
variables $s_1$, $s_2$, $t_1$ and $t_2$, as exhibited in eq.~V-8.8 of
Ref.~\cite{byckling}.  In particular, it will be convenient to express
$t_2$ in terms of  $s_1$, $s_2$, $t_1$ and $\cos\phi$ following
eq.~V-8.9 of Ref.~\cite{byckling},
\beq \label{t2phi}
m_b^2+m_3^2-t_2=\frac{D-2\left[G(s,t_1,s_2,m_a^2,m_b^2,m_1^2)
G(s_1,s_2,s,m_2^2,m_1^2,m_3^2)\right]^{1/2}\cos\phi}{\lambda(s,s_2,m_1^2)}\,,
\eeq
where
\beq
D\equiv \det\begin{pmatrix} 2s & \quad s+s_2-m_1^2 & \quad s-s_1+m_3^2
\\ s+s_2-m_1^2 &\quad 2s_2 & \quad s_2-m_2^2+m_3^2 \\
s-m_a^2+m_b^2 & \quad s_2-t_1+m_b^2 & 0\end{pmatrix}\,.
\eeq
Note that the phase space
distribution in the helicity angle is uniform, as the
integration over $\phi$ in \eq{basic} is trivial.  However, because
the matrix element given in \eq{twotwo} depends on $t\equiv t_2$, the
calculation of the partonic cross section 
for $a+b\to 1+2+3$ will require a nontrivial
integration over $\phi$.

The step functions in \eq{basic} determine the
kinematical ranges of the parameters $s_1$, $s_2$ and $t_1$.
Taking $m_a=m_b=m_1=0$ and $m_2=m_3=m$, it follows that:
\beq \label{dr}
\frac{dR_3}{ds_2 dt_1 ds_1}=\frac{\pi}{8s(s-s_2)}
\,\Theta\{-G(s,t_1,s_2,0,0,0)\}
\Theta\{-G(s_1,s_2,s,m^2,0,m^2)\}\int_0^{2\pi} d\phi\,.
\eeq
The differential cross-section is given by:
\beq
d\sigma=\frac{1}{64\pi^5 s}dR_3(s)|\mathcal{M}(a+b\to 1+2+3)|^2\,,
\eeq
where the squared matrix element is suitably averaged over
initial spins and summed over final spins.  The dominant contribution
to $a+b\to 1+2+3$ takes place via $a+b\to c+3$, where $c$ is
produced approximately on-shell and subsequently decays via
$c\to 1+2$.  In particular, since $c$ is a spin-zero particle,
\beq
|\mathcal{M}(a+b\to 1+2+3)|^2\simeq\frac{|\mathcal{M}(a+b\to c+3)|^2\,
|\mathcal{M}(c\to 1+2)|^2}{(s_1-M^2)^2+M^2\Gamma^2}\,.
\eeq
We now use \eqs{twotwo}{decay} and
employ the narrow width approximation,
\beq
\frac{1}{(s_1-M^2)^2+M^2\Gamma^2}\longrightarrow \frac{\pi}{M\Gamma}
\delta(s_1-M^2)\,.
\eeq
Hence, it follows that:
\beqa \label{sig3}
\frac{d\sigma}{ds_2 dt_1 ds_1}&=&\frac{B}{32\pi^2 \xi s^2(s-s_2)}
\delta(s_1-M^2)  \Theta\{-G(s,t_1,s_2,0,0,0)\} \nonumber \\[6pt]
&&\qquad\,\,\,\times
\Theta\{-G(s_1,s_2,s,m^2,0,m^2)\}\int_0^{2\pi}|C_1(s,t_2)|^2 d\phi \,,
\eeqa
where $t_2$ should be expressed in terms of $s_1$, $s_2$, $t_1$ and 
$\cos\phi$ using \eq{t2phi} before performing the integration 
over $\phi$, and
\beq \label{xidef}
\xi\equiv 1-\frac{m^2}{M^2}\,.
\eeq

Assuming that $G(s_1,s_2,s,m^2,0,m^2)<0$,
we can immediately use the $\delta$-function to integrate over $s_1$.
Using \eq{G}, we obtain:
\beqa
G(s_1,s_2,s,m^2,0,m^2)&=&s_1^2 s_2-s_1 s_2(s-s_2+2m^2)+m^2s(s-s_2)+s_2 m^4
\nonumber \\
&=&s_2(s_1-s_1^+)(s_1-s_1^-)\,,\label{G1}
\eeqa
where $s_2$ is strictly non-negative and
\beq \label{s1cond}
s_1^{\pm}=m^2+\half(s-s_2)\left[1\pm\sqrt{1-\frac{4m^2}{s_2}}\right]\,.
\eeq
That is, we require that:
\beq \label{ineq}
s_1^-\leq M^2\leq s_1^+\,,
\eeq
otherwise, $s_1=M^2$ can never be satisfied when $G(s_1,s_2,s,m^2,0,m^2)<0$.
Note that \eq{ineq} yields upper and lower
limits for $s_2$.  
One can then use \eq{s2} to
obtain upper and lower limits for $E_1$.  These limits correspond to
the roots of the quadratic equation,
\beq
4\sqrt{s} M^2E_1^2-2(M^2-m^2)(s+M^2-m^2)E_1+\sqrt{s}(M^2-m^2)^2=0\,.
\eeq
These roots can be expressed as:\footnote{Note that \eq{e1pm}
is equivalent to $E_1^{\pm}=\half\xi(E_c\pm p_c)$, where
$E_c$ and $p_c$ are the center-of-mass energy and momentum of
the decaying particle $c$.}
\beq \label{e1pm}
E_1^{\pm}\equiv \frac{\xi}{4\sqrt{s}}\left[s+M^2-m^2\pm
\lambda^{1/2}(s,M^2,m^2)\right]\,,
\eeq
where $\xi$ is defined in \eq{xidef}.
Likewise, employing \eq{s2}, we define
\beq
s_2^\pm=s-2\sqrt{s}E_1^{\mp}\,.
\eeq
The range of $t_1$ is determined from the inequality:
\beq
G(s,t_1,s_2,0,0,0)\equiv st_1(s+t_1-s_2)\leq 0\,,\label{range1}
\eeq
where we have used \eq{G} to evaluate the $G$-function.
That is, as $s_2$ ranges over $s_2^-\leq s_2\leq s_2^+$,
\beq
s_2-s\leq t_1\leq 0\,.
\eeq

Assuming that $s_2^-\leq s_2\leq s_2^+$, the integration of
\eq{sig3} over $s_1$ is immediate, and we obtain:
\beq \label{sig2}
\frac{d\sigma}{ds_2 dt_1 }=\frac{B}{32\pi^2\xi s^2(s-s_2)}
\,\Theta\{-G(s,t_1,s_2,0,0,0)\}\int_0^{2\pi}|C_1(s,A_1+A_2\cos\phi)|^2 d\phi  \,,
\eeq
where \eq{t2phi} has been used to write $t_2=A_1+A_2\cos\phi$.
Using \eqs{G1}{range1} with $s_1=M^2$, $m_a=m_b=m_1=0$ and $m_2=m_3=m$, 
the coefficients $A_1$ and $A_2$ are given by
\beqa
A_1&= & m^2-\frac{s_2(s-s_2+t_1)(M^2-m^2)-st_1(s-s_2-M^2+m^2)}{(s-s_2)^2}\,,
\label{Aphi}\\
A_2&= &\frac{2\left[s t_1(s-s_2+t_1)(M^4 s_2-M^2 s_2(s-s_2+2m^2)
+m^2s(s-s_2)+s_2 m^4)\right]^{1/2}}{(s-s_2)^2}\,.\nonumber \\
&&\phantom{line}\label{Bphi}
\eeqa

We now introduce the transverse momentum, $p_T$ of particle 1, which
is defined by $p_T=E_1\sin\theta$.  Note that
\beq \label{G2}
s t_1(s+t_1-s_2)=-s^2 p_T^2\,,
\eeq
which is strictly non-positive as required by \eq{range1}.  In particular,  
\beq \label{sign}
\cos\theta=\pm\sqrt{1-\frac{p_T^2}{E_1^2}}\,,
\eeq
where the $\pm$ indicates that $\theta$ and $\pi-\theta$ correspond
to the same value of $p_T$.  Thus, \eq{t1} yields
\beq \label{t1pt}
t_1=-\sqrt{s}\left(E_1\mp\sqrt{E_1^2-p_T^2}\right)\,.
\eeq

One can now perform a change of variables from $\{t_1\,,\,s_2\}$ to
$\{p_T^2\,,\,E_1\}$.  Computing the Jacobian of the transformation, it
follows that:
\beq \label{dtds}
dt_1 ds_2=\frac{sdp_T^2 dE_1}{\sqrt{E_1^2-p_T^2}}\,.
\eeq
The limits of the kinematic variables $p_T$ and $E_1$ are given by:
\beq \label{range2}
0\leq p_T\leq E_1\,,\qquad\quad E_1^-\leq E_1\leq E_1^+\,,
\eeq
where the range of $p_T$ follows from $|\cos\theta|\leq 1$ and
$E_1^\pm$ is defined in \eq{e1pm}.
Since we aim to compute $d\sigma/dp_T$, it is more useful to interchange
the order of integration.  Thus, equivalent to \eq{range2} is:
\beqa
&&\text{for}~~0\leq p_T\leq E_1^-\,,\,\qquad\qquad
E_1^- \leq E_1\leq E_1^+\,, \label{range3} \\
&&\text{for}~~E_1^-\leq p_T\leq E_1^+\,,\,\qquad\quad
\,\,p_T\leq E_1\leq E_1^+ \,. \label{range4}
\eeqa
Combining \eqs{sig2}{dtds}, and adding the contributions
from the two possible values of $t_1$ that
yield the same value of $p_T$ [cf.~\eq{t1pt}], one obtains:
\beq
\frac{d\sigma}{dp_T^2 dE_1}=\frac{B}
{64\pi^2\xi s^{3/2}E_1\sqrt{E_1^2-p_T^2}}\int_0^{2\pi} d\phi\,
\sum_{j=\pm}|C_1(s,A_1^{(j)}\!+A_2\cos\phi)|^2\,,
\eeq
where $A_1^{(\pm)}$ is defined by \eq{Aphi},
and the superscript indicates which sign is used in \eq{t1pt} to express $t_1$
in terms of $E_1$ and~$p_T^2$.  In contrast, 
$A_2$ [defined in \eq{Bphi}] does not depend on the
sign choice in \eq{t1pt} as a consequence of \eq{G2}.

We now integrate over $E_1$, employing the limits of integration given
in \eqs{range3}{range4}.  Writing $dp_T^2=2p_T dp_T$, we arrive at our final
result,\footnote{There is no singularity in the limit of $\xi\to 0$
since in this limit, $E^\pm\to 0$.}
 \beq \label{bx}
\frac{d\sigma}{dp_T}=\frac{B p_T}{32\pi^2\xi s^{3/2}}
\int_{E_{\rm min}}^{E_{\rm max}}
\frac{dE_1}{E_1\sqrt{E_1^2-p_T^2}}\int_0^{2\pi} d\phi\,
\sum_{j=\pm}|C_1(s,A_1^{(j)}\!+A_2\cos\phi)|^2\,,
\eeq
where the upper and lower limits of
integration are given by $E_{\rm max}\equiv E_1^+$ and
\beq
E_{\rm min}=\begin{cases} E_1^- & \quad \text{for}~~0\leq p_T\leq E_1^{-}\,,
  \\ p_T & \quad \text{for}~~E_1^-\leq p_T\leq E_1^+\,.
\end{cases}
\eeq

As a warmup, we shall ignore the details of the scattering matrix element
for the process $a+b\to c+3$ by putting $C_1=1$.  In this case,
the integrals in \eq{bx} are elementary, and the end result is:
\beq \label{sigpT}
\frac{d\sigma}{dp_T}=\frac{B}{8\pi\xi s^{3/2}}\left[
\tan^{-1}\left(\frac{\sqrt{[E_1^+]^2-p_T^2}}
{p_T}\right)-\Theta(E_1^--p_T)\tan^{-1}\left(\frac{\sqrt{[E_1^-]^2-p_T^2}}
{p_T}\right)\right]\,,
\eeq
where $0\leq p_T\leq E_1^+$, and the step function $\Theta$ is defined as usual,
\beq
\Theta(E_1^--p_T)=\begin{cases} 1 &
\quad \text{for}~~0\leq p_T\leq E_1^{-}\,, \\ 0 &
\quad \text{for}~~E_1^-\leq p_T\leq E_1^+\,.
\end{cases}
\eeq

It is convenient to introduce dimensionless variables,
\beq \label{dimless}
w\equiv \frac{2E_1}{\sqrt{s}}\,,\qquad\quad
x\equiv \frac{2p_T}{\sqrt{s}}\,,\qquad\quad y\equiv\frac{M^2}{s}\,,
\quad\qquad z\equiv 1-\xi=\frac{m^2}{M^2}\,.
\eeq
The kinematics of the scattering process requires that $\sqrt{s}\geq M+m$, which is equivalent to the
condition,
\beq \label{inequal}
\sqrt{y}\,(1+\sqrt{z})\leq 1\,.
\eeq
The range $w$ is given by $w^-\leq
w\leq w^+$, where 
\beq \label{wpmdef}
w^\pm=\half (1-z)\left[1+y(1-z)\pm\lambda^{1/2}(1,y,yz)\right]\,.
\eeq
Hence, the range of $x$ is 
\beq \label{rangex}
0\leq x\leq w^+<1\,.
\eeq
As an example, take $y=0.5$ and $z=0.1$, which is consistent with the inequality given
in \eq{inequal}.  \Eq{rangex} then implies that $0\leq x\leq 0.79657$.
The transverse momentum distribution, plotted in Fig.~\ref{jacob}
exhibits a striking Jacobian peak located at $x=0.50843$, which
corresponds to 
\beq \label{ptpeak}
(p_T)_{\rm peak}=E_1^-\,.
\eeq

\FIGURE[t]{
\epsfig{file=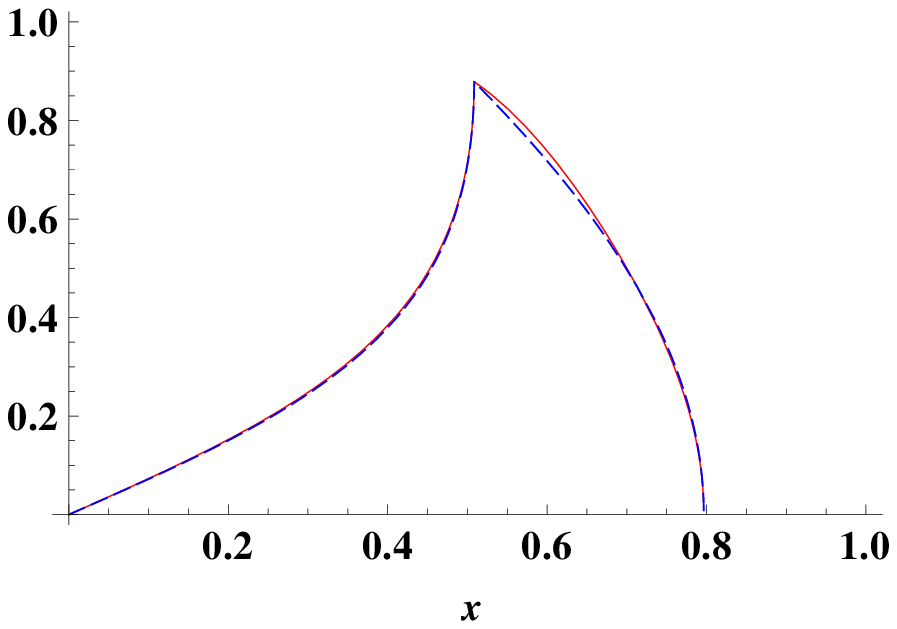,width=4in}
\caption{\label{jacob} Unnormalized $p_T$ distributions for
$a+b\to c+3\,,\,c\to 1+2$, assuming that the matrix element 
for $a+b\to c+3$ is
constant (dashed curve) or is given by \eq{sqme} (solid curve).  The
rescaled transverse momentum is defined by $x\equiv 2p_T/\sqrt{s}$ and can
take on values
in the range $0\leq x\leq x_{\rm max}$, where $x_{\rm max}\equiv\frac{1}{2}(1-z)\left[1+y(1-z)+\lambda^{1/2}(1,y,yz)\right]$.
The masses of particles $c$ and $d$ are fixed by $y\equiv M^2/s=0.5$
and $z\equiv m^2/M^2=0.1$.
To facilitate the comparison of the two $p_T$ distributions,
the relative normalization of the two curves has been fixed
such that the height of the peaks of the distributions coincide.  The
\textit{location} of the peak at $x=0.50843$,
corresponding to \eq{ptpeak}, is the same for both curves.}
}

The origin of the Jacobian peak is a consequence of the change in
kinematic variables given in \eq{dtds}, and is rather insensitive to
the form of the matrix element.  To illustrate this point, we
have numerically evaluated \eq{bx}, where the tree-level form for
$C_1$ for $gq\to \widetilde{q}_R {\widetilde \chi}_1^0$ is
employed \cite{Haber:1984fi,Dawson:1983fw,matel}
\beq \label{sqme}
|C_1(s,t)|^2=N\left[\frac{s+t-M^2}{2s}-\frac{M^2(m^2-t)}{(M^2-t)^2}
+\frac{sm^2+(m^2-t)(M^2-m^2)}{s(M^2-t)}\right]\,,
\eeq
where $N$ is an overall dimensionless normalization factor that
depends on the relevant couplings.  In terms of the dimensionless
variables introduced in \eq{dimless}, 
\beq
\frac{d\sigma}{dx}=\frac{Bx}{64\pi^2 s(1-z)}\int_{w_{\rm min}}^{w_{\rm max}}
\frac{dw}{w}\frac{1}{\sqrt{w^2-x^2}}\int_0^{2\pi} d\phi \sum_{j=\pm}
|C_1(s, A^{(j)}_1\!+ A_2\cos\phi)|^2\,,
\eeq
where
\beq
w_{\rm max}=w^+\,,\qquad w_{\rm min}=\begin{cases} w^-\,, & \,\,\,
\text{for $0\leq x\leq w^-\,,$} \\ x\,, & \,\,\, \text{for $w^-\leq x\leq w^+\,,$}
\end{cases} 
\eeq
and the coefficients $A_1^{(\pm)}$ and $A_2$ are given by:
\beqa
A^{(\pm)}_1&=&\half s\biggl\{y(1+z)-1\pm 
\frac{\sqrt{w^2-x^2}\left[w+y(1-z)(w-2)\right]}{w^2}\biggr\} \,,\\
A_2&=& \frac{sxy^{1/2}}{w^2}\left[w(1-w-z)-y(1-w)(1-z)^2\right]^{1/2}\,.
\eeqa

The resulting unnormalized
$p_T$ distribution is exhibited in Fig.~\ref{jacob}.  Note
that the shape of the $p_T$ distribution
is dominated by the explicit kinematic factors that
appear in \eq{bx}, and depends quite weakly on the actual form of the
squared-matrix element given in \eq{sqme}.  Moreover, the location of
the peak in the $p_T$ distribution is unchanged and given by
\eq{ptpeak}, as a consequence of structure of the kinematic limits
given in \eqs{range3}{range4}.

In the above analysis, the location of the Jacobian peak given in
\eq{ptpeak} depends on 
the partonic center-of-mass
energy $\sqrt{s}$. 
The differential cross section for the hadronic scattering process,
$A+B\to c+3+X\to 1+2+3+X$, is obtained by convoluting the $p_T$
distribution of the partonic subprocess, $a+b\to c+3\to 1+2+3$,
with the product of the parton distribution functions
$f^A_a(x_1,Q^2)f^B_b(x_2,Q^2)$, where the total center-of-mass
squared-energy $S$ is related to the partonic center-of-mass energy
via $s=x_1 x_2 S$, and $Q$ is the factorization scale.  In the convolution,
partonic center-of-mass energies close to the
energy threshold for the partonic process provide the dominant contribution
to the production of the final state.
In this case, one can derive an approximate formula for the location
of the Jacobian peak that does
not depend on the partonic center of mass energy.  The threshold
for $a+b\to c+3$ corresponds to the point at which
\beq
\lambda(s,M^2,m^2)=(s+M^2-m^2)^2-4sM^2=0\,.
\eeq
At this point $s+M^2-m^2=2M\sqrt{s}$ (or equivalently, $\sqrt{s}=M+m$), in which case
\beq \label{e1m}
E_1^-= E_1^+=\frac{M^2-m^2}{2M}\,.
\eeq
Of course, the cross-section given in \eq{bx} vanishes
exactly at threshold where $E_1^-=E_1^+$.  
However, if we are close to threshold, then
\eq{e1m} still provides a decent approximation to $E_1^-$, in which
case the location of the Jacobian peak is:
\beq \label{peak}
(p_T)_{\rm peak}=E_1^-\simeq \frac{M^2-m^2}{2M}=\half\xi M\,,
\eeq
which is independent of the partonic center-of-mass energy.

In this paper, we have numerically computed the transverse momentum
distribution of the hadronic scattering process, taking into account
the partonic scattering process at all allowed values of the partonic
center-of-mass energy.  In particular, as the partonic
center-of-mass energy is increased above
the threshold energy for $a+b\to c+3$, the location of the peak
of the partonic transverse momentum distribution, $E_1^{-}$ [cf.~\eq{e1pm}]
\textit{decreases} relative to the estimate given in \eq{peak}.
Thus, we expect the actual peak in the transverse momentum
distribution of the hadronic scattering process (or equivalently in the
missing transverse energy distribution) to be somewhat less
than the result of \eq{peak}.  This is indeed the case in the
$\ptmiss$ distributions that we exhibit in this paper.

Note that in the approximation that
the transverse momentum of particle $c$ is due entirely from the hard scattering
process (i.e.~the transverse momentum of the initial partons and the spectators
are neglected), the distribution of the missing transverse energy
(i.e.~particles 2 and 3 of the hard scattering process) should
precisely match that of the transverse momentum of the monojet (i.e.~particle 1
of the hard scattering process).  Of course, the effects of spectators,
initial and final state radiation, fragmentation of final state partons,
jet mismeasurements and detector effects
will tend to reduce the sharpness of the peak in the $\ptmiss$
distributions as compared to that of Fig.~\ref{jacob}.

\end{document}